\documentclass[12pt]{article}
\usepackage[latin1]{inputenc}
\usepackage[T1]{fontenc}
\usepackage[svgnames]{xcolor}
\usepackage[pdftex]{graphicx}
\usepackage{floatflt}
\usepackage{bibentry}
\usepackage[nottoc]{tocbibind}
\usepackage[square,numbers,sort&compress]{natbib}
\RequirePackage[pdftex,pdfpagemode=none, pdftoolbar=true,
pdffitwindow=true,pdfcenterwindow=true]{hyperref}
\DeclareGraphicsExtensions{.pdf,.eps,.png,.jpg}
\usepackage{color}
\usepackage{amsmath}
\usepackage{amsfonts}
\usepackage{amssymb}
\usepackage{graphicx}
\usepackage{pgfplots}
\usepackage{tikz}

 \newcommand{\bs}{\boldsymbol}
\newcommand{\Rb}{\ensuremath{^{87}}Rb }

\newcommand{\hpsi}{\ensuremath{\hat{\psi}}}

\newcommand{\lT}{\ensuremath{\lambda_{\rm T}}}

\newcommand{\um} {\ensuremath{\mu \rm m} }
\newcommand{\uK}{ \ensuremath{\mu \rm K} }
\newcommand{\kB}{\ensuremath{k_{ \rm B}} }

\newcommand{\BECp}{\ensuremath{\rm{BEC}_{\perp}}}
\newcommand{\BECf}{\ensuremath{\rm{BEC}_{\rm{full}}}}
 \newcommand{\tev}{t_{\rm evap}}
 \newcommand{\tho}{t_{\rm hold}}
 \newcommand{\Nv}{N_{\rm v}}
  \newcommand{\vj}{{\bs j}}

 \newcommand{\D}{{\rm d}}
 \newcommand{\E}{{\rm e}}
  \newcommand{\I}{{\rm i}}
 \newcommand{\ZZ}{\mathbb{Z}}   
  
\title{Emergence of coherence in \\ a uniform quasi-two-dimensional Bose gas}
\date{\today}
\author{Lauriane Chomaz$^{1}$, Laura Corman$^{1}$, Tom Bienaimé$^1$, Rémi Desbuquois$^2$,\\ Christof Weitenberg$^3$, Sylvain Nascimbène$^1$, Jérôme Beugnon$^{1}$, Jean Dalibard$^{1}$ }

\begin{document}
\maketitle
\begin{center}
\textsl{\small $^1$Laboratoire Kastler Brossel, Collège de France, ENS, CNRS, UPMC,}\par
\textsl{\small 11 Place Marcelin Berthelot, 75005 Paris, France}\par
\textsl{\small $^2$Institut für Quantenelektronics, ETH Zurich, 8093 Zurich, Switzerland}\par
\textsl{\small $^3$Institut für Laserphysik, Universität Hamburg, Luruper Chaussee 149, D-22761 Hamburg, Germany}\par
\end{center}
\vspace{2cm}

\begin {abstract}
Phase transitions are ubiquitous in our three-dimensional world. By contrast most conventional transitions do not occur in infinite uniform two-dimen\-sional systems because of the increased role of thermal fluctuations. Here we explore the dimensional crossover of Bose--Einstein condensation  (BEC) for a weakly interacting atomic gas confined in a novel quasi-two-dimensional geometry, with a flat in-plane trap bottom. We detect the onset of an extended phase coherence, using  velocity distribution measurements and matter-wave interferometry. We relate this coherence to the transverse condensation phenomenon, in which a significant fraction of atoms accumulate in the ground state of the motion perpendicular to the atom plane. We also investigate the dynamical aspects of the transition through the detection of topological defects that are nucleated in a quench cooling of the gas, and we compare our results to the predictions of the Kibble--Zurek theory for the conventional BEC second-order phase transition. 
\end {abstract}

\vskip 1cm

Bose--Einstein condensation (BEC) is a remarkably simple phase transition that can in principle occur in a fluid even in the absence of interatomic interactions. As a mere result of single-particle statistics, a phase-coherent fraction appears in the fluid, described by a uniform wave-function spanning the whole system. During the last two decades, cold atom experiments have been used to probe many aspects of BEC \cite{Pethick:2002,Pitaevskii:2003,Leggett:2006}. However, most of these cold atom studies are performed in the presence of a harmonic confinement. BEC becomes in this case a local transition: the condensate forms at the center of the trap where the density is the largest, and interactions between particles play a dominant role in the equilibrium state of the fluid. In this geometry the non-homogeneous character of the gas makes it difficult to address some important features of BEC, such as the existence of long-range phase coherence.  The recent achievement of a three-dimensional (3D) Bose gas undergoing BEC in a box-like potential \cite{Gaunt:2013,Gotlibovych:2014} constitutes an important step forward,  realizing the text-book paradigm of an extended and uniform coherent matter wave.

When turning to low-dimensional  (low-D) systems, subtle effects emerge due to the entangled roles of Bose statistics and thermal fluctuations.  
First, in an infinite low-D\  ideal gas, no BEC is expected at non-zero temperature, because of the modification of the single-particle density of states with respect to the 3D case \cite{Huang:1987}. In other words, the phase coherence between two points tends to zero when their distance increases, contrary to the 3D situation. Second, Bose statistics may facilitate the  freezing of some directions of space required to produce a  low-D system.  Consider the uniform two-dimensional (2D) case obtained by imposing a tight harmonic trapping potential (frequency $\nu_z$) along the third direction $z$. The transverse condensation phenomenon \cite{vand97,Armijo:2011,RuGway:2013} allows one to reach an effective 2D situation even in the ``thermally unfrozen'' regime, where the quantum $h\nu_z$ is smaller than the thermal energy $\kB T$ ($h$ and $\kB$ stand for Planck's and Boltzmann's constants). Third,  for the 2D case\ in the presence of interactions between the particles, the situation gets more involved with the possibility of a superfluid, Berezinskii--Kosterlitz--Thouless (BKT) transition \cite{Berezinskii:1971,Kosterlitz:1973} for a large enough phase-space-density, even though the absence of true long-range coherence remains valid \cite{Mermin:1966,Hohenberg:1967}. This superfluid transition has been identified and characterized over the recent years with non-homogeneous, harmonically trapped Bose gases \cite{Hadzibabic:2006,Clade:2009,Tung:2010,Hung:2011,Desbuquois:2012,Ha:2013,Choi:2013b}.

Another key feature of phase transitions for uniform systems is the time needed to establish the coherence/quasi-coherence over the whole sample.  
As is well known for critical phenomena  \cite{Hohenberg:1977}, the coherence length and the thermalization time diverge at the transition point, thus limiting the size of the phase-coherent domains that are formed at its crossing. The Kibble--Zurek (KZ) theory \cite{Kibble:1976,Zurek:1985} allows one to evaluate the scaling of the domain size with  the speed of the crossing. 
Once the transition has occurred, these domains start merging together. During this coarsening dynamics singularities  taking the form of topological defects can  be nucleated at their boundaries, with a spatial density  directly related to the characteristic domain size.  The KZ mechanism has been studied in a variety of experimental systems (see for example \cite{Chuang:1991,Ruutu:1996,Bauerle:1996,Ulm:2013,Pyka:2013,Monaco:2009}), including cold atomic gases \cite{Sadler:2006,Weiler:2008,Chen:2011,Lamporesi:2013,Braun:2014,Corman:2014,Hadzibabic:2014}.  In 2D quantum fluids, the singularities  take the form of quantized vortices, \emph{i.e.}, points of zero density around which   the macroscopic wavefunction of the gas has a $\pm 2\pi$ phase winding. 

In this paper we present an experimental realization  of a uniform atomic Bose gas  in a quasi-2D geometry, addressing both the steady state of the fluid and  its quench dynamics. First with the gas in thermal equilibrium,  we characterize the threshold for the emergence of an extended  phase coherence by two independent methods, based  on (i) the measurement of the atomic velocity distribution  and (ii) matter-wave interferences. We show in particular that for the thermally unfrozen case $\zeta \gg 1$ (with $\zeta=\kB T / h \nu_z$), the transverse condensation phenomenon induces an extended in-plane coherence. 
 Second, we explore the quench dynamics of the gas prepared in an initial state such that $\zeta \gg 1$ and observe density holes associated to vortices. We study the relation between the cooling rate  and the number of vortices that subsist after a given relaxation time, and we compare our results with the predictions of the Kibble--Zurek theory. 

\begin{figure}
\begin{center}
\begin{minipage}[c]{11cm}
\includegraphics[width=11cm]{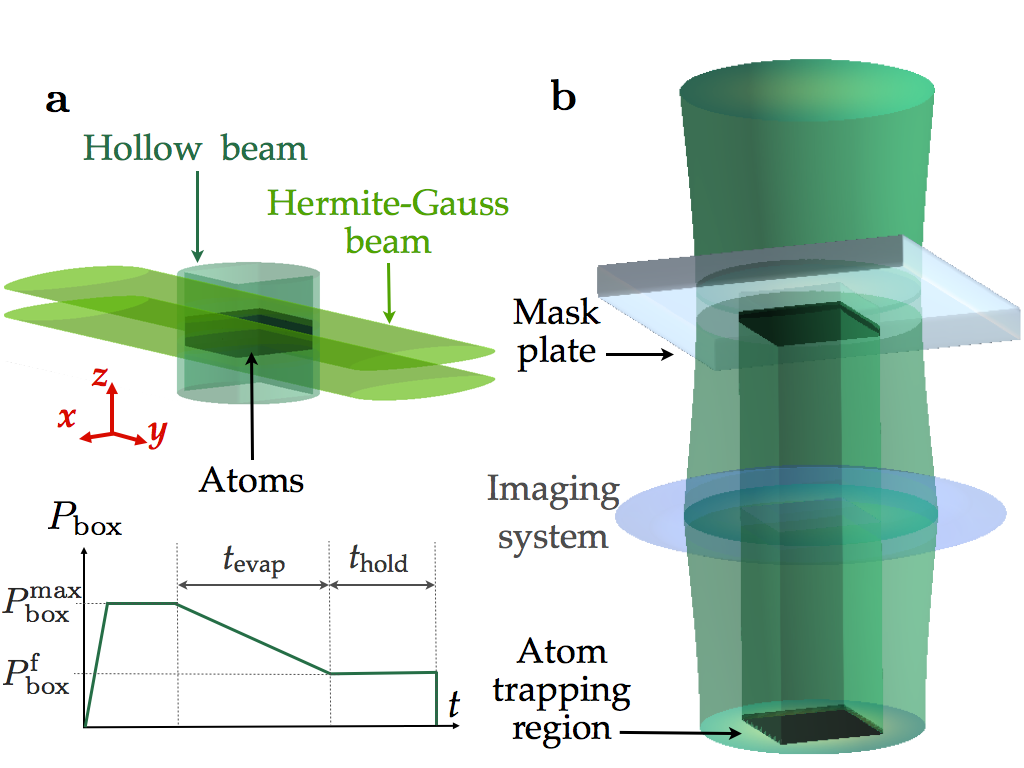} 
\end{minipage}
\begin{minipage}[c]{3.6cm}
\includegraphics[trim=350 0 350 0, clip=true, width=3.6cm]{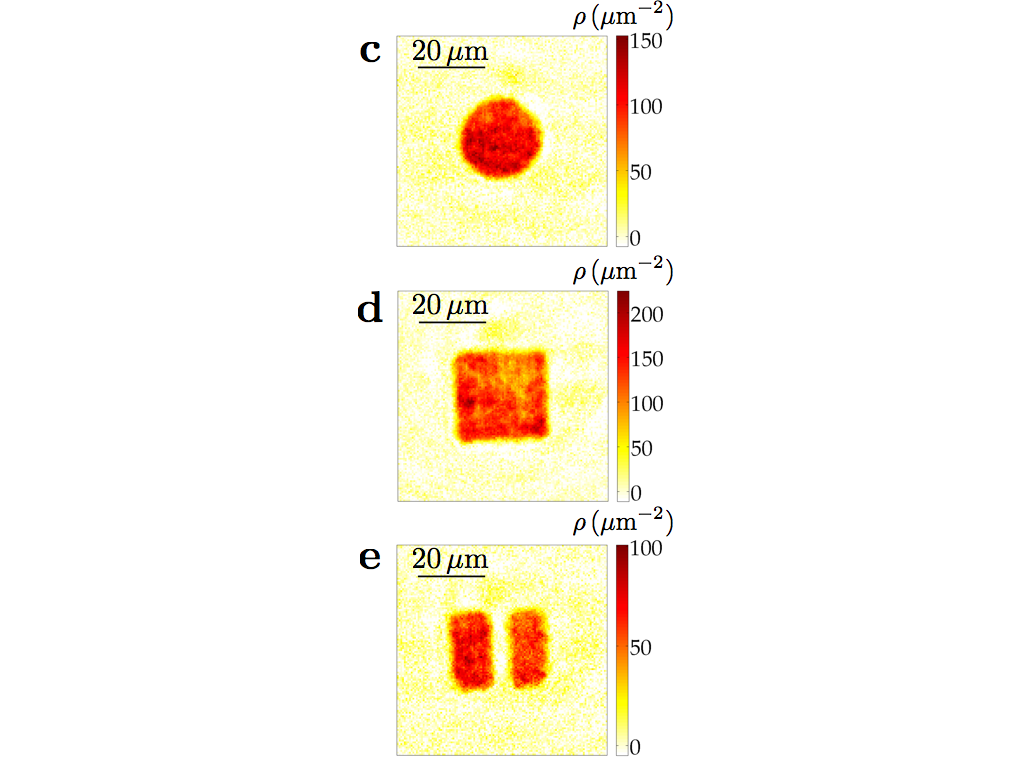} 
\end{minipage}
\end{center}
\caption{ \sl \textbf{ Production of uniform Bose gases  in quasi-2D geometries.} \textbf{(a)}  We slice a horizontal sheet from a 3D cold gas of\ \ \Rb atoms using a blue detuned laser beam propagating along $x$ and shaped with an intensity node in the $z=0$ plane. It creates an adjustable harmonic confinement along $z$ of frequency $\nu_z = 350$--$1500\,$Hz. We superimpose a hollow beam  propagating along $z$ and producing a uniform confinement in the $xy$ plane (see \textbf{b}). The power $P_{\rm box}$ of this beam is ramped up in $10\,$ms to its maximal value $P_{\rm box}^{\rm max}$ corresponding to a potential barrier $U_{\rm box} \sim \kB × 3\, \uK$ for the \Rb atoms. After holding $P_{\rm box}$ at $P_{\rm box}^{\rm max}$ for $0.5\,$s, we lower it linearly to its final value $P_{\rm box}^{\rm f}$  in a typical time of $t_{\rm evap}=2\,$s and keep it constant for a typical $t_{\rm hold} = 0.5\,$s. We vary $P_{\rm box}^{\rm f}$ to adjust the final temperature of the gas via evaporative cooling. \textbf{(b)}  The in-plane ($xy$) confinement is provided by a blue-detuned laser beam shaped by placing a dark intensity mask on its path and imaging it at the position of the atoms. \textbf{(c, d, e)}: In-situ density distributions of uniform gases trapped in a disk of radius $R=12\,\um$, a square box of length $L=30\,\um$, and two coplanar and parallel rectangular boxes of size $24\times 12\,\um^2$, spaced by $d=4.5\,\um$. These distributions are imaged using a high intensity absorption imaging technique (see methods). }
\label{fig:setup}
\end{figure}

\section*{Results}
\noindent{\textbf{Production of uniform gases in quasi-2D geometries.}} 
We  prepare a cold  3D gas of rubidium ($^{87}$Rb) atoms using standard laser and evaporative cooling techniques. Then we transfer the gas in a trap formed with two orthogonal laser beams at  wavelength 532\,nm, shorter than the atomic resonance wavelength (780\,nm), so that the atoms are attracted towards the regions of low light intensity {(Fig.\;\ref{fig:setup}\,\textbf{a})}. The strong confinement along the $z$ direction (vertical) is provided by a  laser beam propagating along the $x$ direction. It is prepared in a Hermite--Gauss mode, with a node in the plane $z=0$, and  provides a harmonic confinement along the $z$ direction with a frequency $\nu_z$ in the range $350-1500$\,Hz. For the confinement in the horizontal $xy$ plane, we realize a box-like potential by placing an intensity mask on the second laser beam path, propagating along the $z$ direction {(Fig.\;\ref{fig:setup}\,\textbf{b})}. Depending on the study to be performed, we can vary the shape (disk, square, double rectangle) and the area  ${\cal A}$ (from 200 to 900\,$\mu$m$^2$) of the region accessible to the gas in the plane. The relevance of our system for the study of 2D physics is ensured by the fact that the size of the ground state along the $z$ direction $a_z = \sqrt{h / (m \nu_z)}/2\pi \sim 0.3$ -- $0.6\,\um$ is very small compared to the in-plane extension  $\sqrt{\cal A} \sim 15$ -- $30\,\um$.   The number of atoms $N$ that can be stored and reliably detected in this trap ranges between $1000$ and $100\,000$. We adjust  the temperature of the gas in the interval $T\sim 10 - 250$\,nK by varying the intensity of the beam creating the box potential, taking advantage of evaporative cooling on the edges of this box. The ranges spanned by $\nu_z$ and $T$ allow us to explore the dimensional crossover between the thermally frozen regime ($\zeta\ll 1$) and the unfrozen one ($\zeta\gg 1$). Examples of in situ images of 2D gases are shown in Fig.\;\ref{fig:setup}\,\textbf{c,d,e}. 

\vskip 5mm
\noindent{\textbf{Phase coherence in 2D geometries.}}
For an ideal gas an important consequence of Bose--Einstein statistics is to increase the range of phase coherence with respect to the prediction of  Boltzmann statistics. Here coherence is  characterized  by the one-body correlation function $G_1(\bs r)=\langle \hat{\psi}^\dagger (\bs r) \,\hat{\psi}(0)\rangle$, where $\hpsi(\bs r)$ [resp. $\hpsi^\dagger(\bs r)$] annihilates (resp. creates) a particle in $\bs r$, and where the average is taken over the equilibrium state at temperature $T$. For a gas of particles of mass $m$ described by Boltzmann statistics, $G_1(r)$ is a Gaussian function $\propto \exp(-\pi r^2/\lT^2)$, where $\lT=h/(2\pi m\kB T)^{1/2}$ is the thermal wavelength. 

Consider the particular case of a 2D  Bose gas (\emph{e.g.}, $\zeta\ll 1$). When its phase-space-density ${\cal D}\equiv \rho \lT^2$ becomes significantly larger than 1 ($\rho$ stands for the 2D spatial density), the structure of $G_1(r)$ changes. In addition to the Gaussian function mentioned above, a broader feature $\propto \exp(-r/\ell)$ develops,  with the characteristic length $\ell$ that increases exponentially with ${\cal D}$ (see \cite{Hadzibabic:2011} and supplementary material)
\begin{equation}
\ell=  \frac{\lT}{\sqrt{4\pi}}\,\exp({\cal D}/2).
\label{eq:ell}
\end{equation}
Usually two main effects amend this simple picture:\begin{itemize}

\item 
In a finite system, when the predicted value of $\ell$ becomes comparable to the size $L$ of the gas, one recovers a standard Bose--Einstein condensate, with a macroscopic occupation of the ground state of the box potential \cite{Petrov:2000a}. The $G_1$ function then takes non-zero values for any $r\leq L$ and the phase coherence extends over the whole area of the gas.

\item In the presence of weak repulsive interactions, the increase of the range of $G_1$ for ${\cal D}\gtrsim 1$ is accompanied with a reduction of density fluctuations, with the formation of a ``quasi-condensate'' or ``pre-superfluid''  state \cite{Prokofev:2002,Clade:2009,Tung:2010}. This state is a medium that can support vortices, which will eventually pair at the superfluid  BKT transition for a larger phase-space-density, around ${\cal D}\sim 8-10$ for the present strength of interactions \cite{Prokofev:2002}. At the transition point, the coherence length $\ell$ diverges and above this point, $G_1(r)$ decays algebraically.
\end{itemize}

\vskip 5mm
\noindent{{\textbf{Role of the third dimension for in-plane phase coherence.}}
When the thermal energy $\kB T$ is not negligibly small compared to the energy quantum  $h \nu_z$ for the tightly confined dimension, the dynamics associated to this direction brings interesting novel features to the in-plane coherence. First, we note that the function $G_1(r)$ can be written in this case as a sum of contributions of the various states $j_z$ of the $z$ motion (see supplementary material). The term with the longest range corresponds to the ground state $j_z=0$, with an expression similar to (\ref{eq:ell}) where ${\cal D}$ is replaced by the phase-space-density ${\cal D}_0$ associated to this state. Now, consider more specifically the unfrozen regime $\zeta \gg1$. In this case one expects that for very dilute gases only a small fraction $f_0$ of the atoms occupies the $j_z=0$ state; Boltzmann statistics indeed leads to $f_0 = 1-e^{-1/\zeta}\approx 1/\zeta \ll 1$. However for large total phase-space densities ${\cal D}_{\rm tot}$, Bose--Einstein statistics modifies this result through the \emph{transverse condensation phenomenon} (\BECp) \cite{vand97}: The phase-space-density that can be stored in the excited states $j_z\neq 0$ is bounded from above, and ${\cal D}_0$ can thus become comparable to ${\cal D}_{\rm tot}$. This large value of ${\cal D}_0$ leads to a fast increase of the corresponding range of $G_1(r)$, thus linking the transverse condensation to an extended coherence in the $xy$ plane. This effect plays a central role in our experimental investigation.

\begin{figure}[p]
\begin{center}
\includegraphics[trim=0 200 0 0, clip=true, width=15cm]{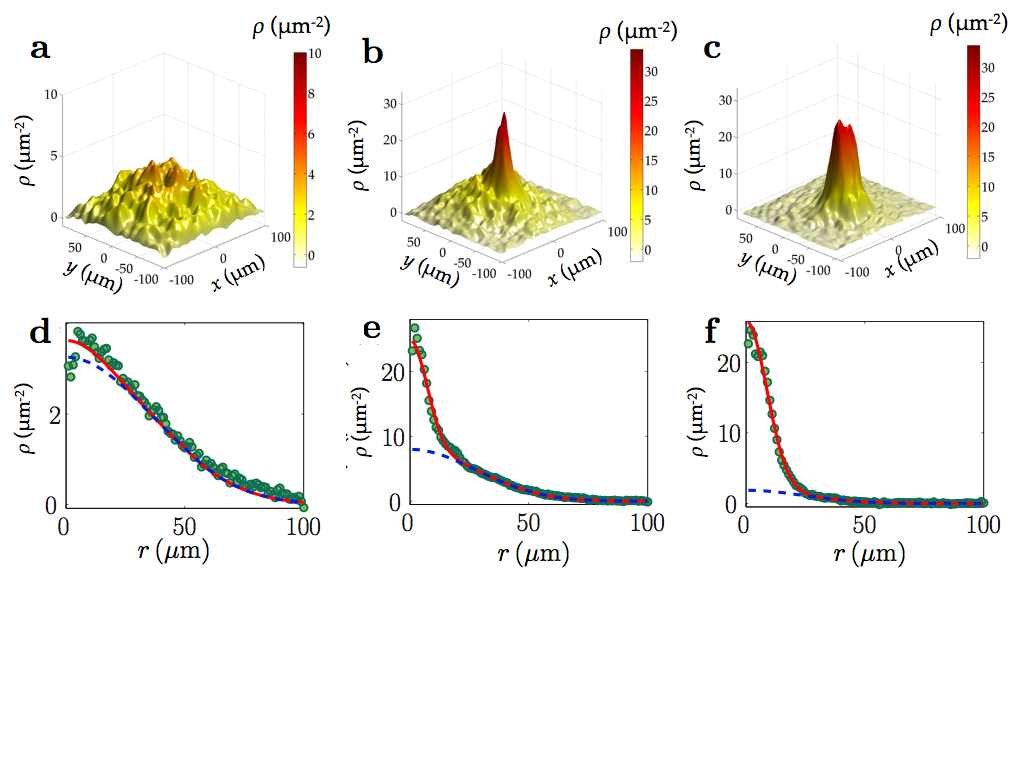}
\end{center}

\begin{center}
\includegraphics[trim=0 10 0 0, clip=true, width=7cm]{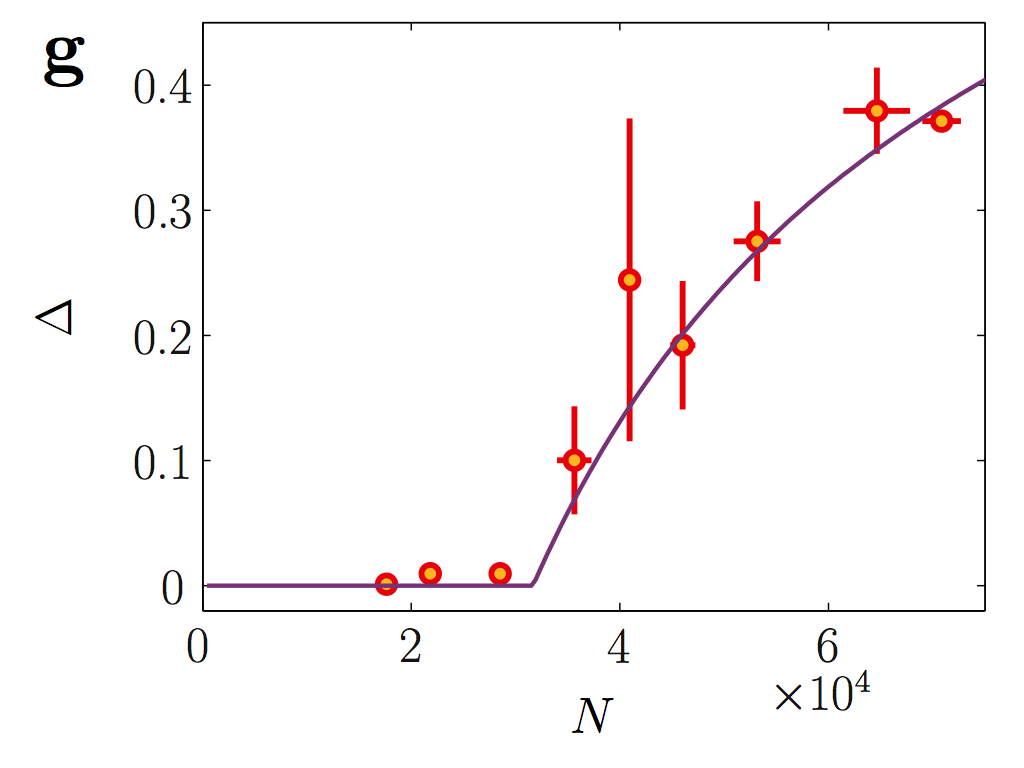} 
\end{center}
\caption{ \sl \textbf{Emergence of bimodal velocity distributions.}
\textbf{(a-f)}\ Surface density distribution $\rho(x,y)$ (first row) and corresponding radial distributions (green symbols) obtained by azimuthal average (second row). The distribution is measured after a $12$\,ms time-of-flight for a gas initially confined in a square of size $L=24\,\um$, with a trapping frequency $\nu_z=365$\,Hz along the $z$ direction.  The temperatures $T$ and atom numbers $N$ for these three realizations are \textbf{a,d}:  (155\,nK, 28\,000), \textbf{b,e}: (155\,nK,\;38\,000), \textbf{c,f}: (31\,nK,\;19\,000). The continuous red lines are fits to the data by a function consisting in the sum of two Gaussians  corresponding to $N_1$ and $N_2$ atoms ($N=N_1+N_2$). The Gaussian of largest width ($N_2$ atoms) is plotted as a blue dashed line. The bimodal parameter $\Delta=N_1/N$  equals \textbf{a,d}: 0.01, \textbf{b,e}: 0.12 and \textbf{c,f}: 0.60.\  \textbf{(g)}  Variation of $\Delta$ with $N$ for a gas in the same initial trapping configuration as \textbf{a-f} and  for $T=155$\,nK (red symbols).\ Error bars are the standard errors of the mean of the binned data set  (with 4 images per point on average). The solid line is a fit to the data by the function $f(N)=(1-(N_{\rm c}/N)^{0.6})$ for $N>N_{\rm c}$, and $f(N)=0$ for $N\leq N_{\rm c}$,  from which we deduce $N_{\rm c}(T)$.  Here  $N_{\rm c}=3.2\,(1)\times 10^4$, where the uncertainty range is obtained by a jackknife resampling method, \emph{i.e.} fitting samples corresponding to a randomly chosen fraction of the global data set. }
\label{fig:ToF}
\end{figure} 

\vskip 5mm
\noindent{{\textbf{Phase coherence revealed by velocity distribution measurements.}}
To characterize the coherence of the gas, we study the velocity distribution, \emph{i.e.}, the Fourier transform of the $G_1(r)$ function. We approach this velocity distribution in the $xy$ plane by performing a 3D time-of-flight (3D ToF):
  We suddenly switch off the trapping potentials along the three directions of space, let the gas expand for a duration $\tau$, and finally image the gas along the $z$ axis.  In such a  3D\ ToF, the gas first expands very fast along the initially strongly confined direction $z$. Thanks to this fast density drop, the interparticle interactions play nearly no role during the ToF and the slower evolution in the $xy$ plane is governed essentially by the initial velocity distribution of the atoms.  The  time-of-flight (ToF)\ duration $\tau$ is chosen so that the size expected for a Boltzmann distribution $ \tau \sqrt{\kB T/m}$ is at least twice the initial extent of the cloud.  Typical examples of ToF images are given in  Fig.\;\ref{fig:ToF}\,\textbf{a-f}. Whereas for the hottest and less dense configurations, the spatial distribution after ToF has a quasi-pure Gaussian shape, a clear non-Gaussian structure appears for larger $N$ or smaller $T$. A sharp peak emerges at the center of the cloud of the ToF picture, signaling an increased occupation of the low-momentum states  with respect to Boltzmann statistics, or equivalently a coherence length significantly larger than $\lT$.

In order to analyze this velocity distribution, we chose as a fit function the sum of two Gaussians of independent sizes and amplitudes, containing $N_1$ and $N_2$ atoms, respectively (see Fig.\;\ref{fig:ToF}\,\textbf{d-f}). We consider the \emph{bimodality parameter} $\Delta=N_1/N$  defined as the ratio of the  number of atoms $N_1$  in the sharpest Gaussian to the total atom number $N=N_1+N_2$.\ 
A typical example for the variations of $\Delta$ with $N$ at a given temperature is shown in  Fig.\;\ref{fig:ToF}\,\textbf{g} for an initial gas with a square shape (side length $L=24\,\um$). It shows a  a sharp crossover, with essentially no bimodality ($\Delta\ll 1$) below a critical atom number $N_{\rm c}(T)$ and a fast increase of $\Delta$ for $N>N_{\rm c}(T)$. We extract the value $N_{\rm c}(T)$ by fitting the function $\Delta \propto (1-(N_{\rm c}/N)^{0.6})$ to the data.  We chose this function as it provides a good representation of the predictions for an ideal Bose gas in similar conditions (see methods).  

\vskip 5mm
\noindent{\textbf{Phase coherence revealed by matter-wave interference.}}
Matter-wave interferences between independent atomic or molecular clouds is a powerful tool to monitor the emergence of extended coherence  \cite{Andrews:1997a,Hadzibabic:2006,Hofferberth:2007,Kohstall:2011,Gaunt:2013}. To observe these interferences in our uniform setup, we first produced two independent gases of similar density and temperature confined in two coplanar parallel rectangles, separated by a distance of $4.5\,\mu$m along the $x$ direction (see Fig.\;\ref{fig:setup}\,\textbf{e}). Then we suddenly released the box potential providing confinement in the $xy$ plane, while keeping the confinement along the $z$ direction (2D ToF). The latter point ensures that the atoms stay in focus with our imaging system, which allows us to observe interference fringes with a good resolution in the region where the two clouds overlap.  A typical interference pattern is shown in Fig.\;\ref{fig:fringes}\,\textbf{a}, where the fringes are (roughly) parallel to the $y$ axis, and show some waviness that is linked to the initial phase fluctuations of the two interfering clouds.

\begin{figure}
\begin{center}
\includegraphics[trim=0 260 0 0,clip=true,width=16cm]{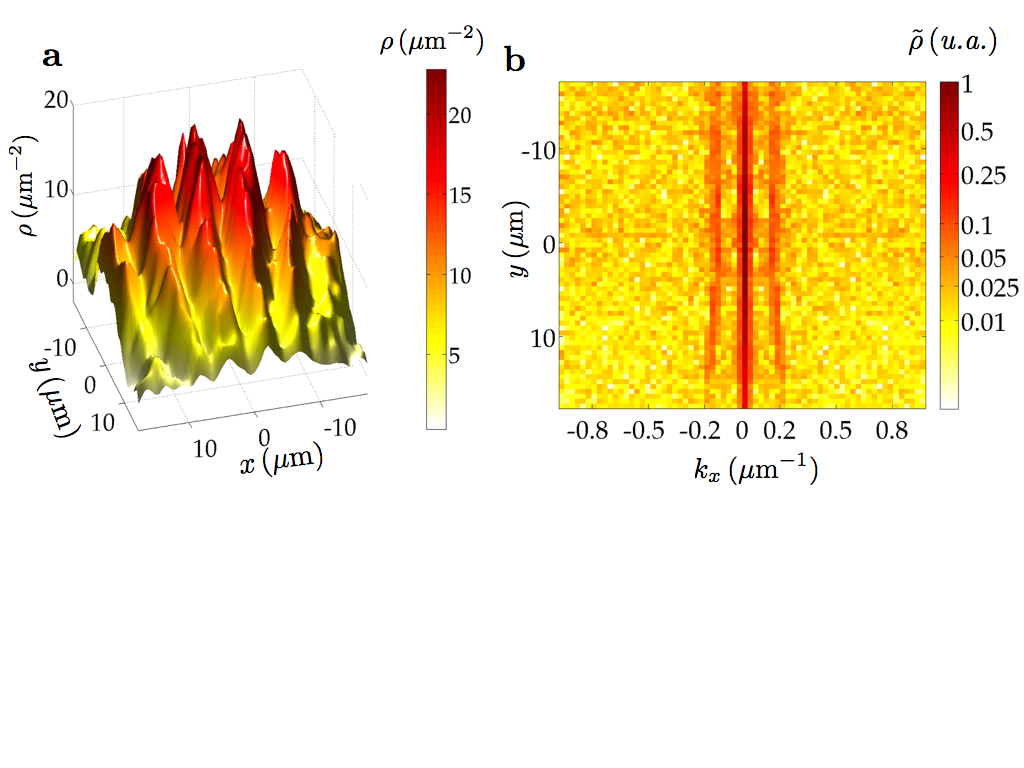}
\includegraphics[trim=0 20 0 20, clip=true,width=9cm]{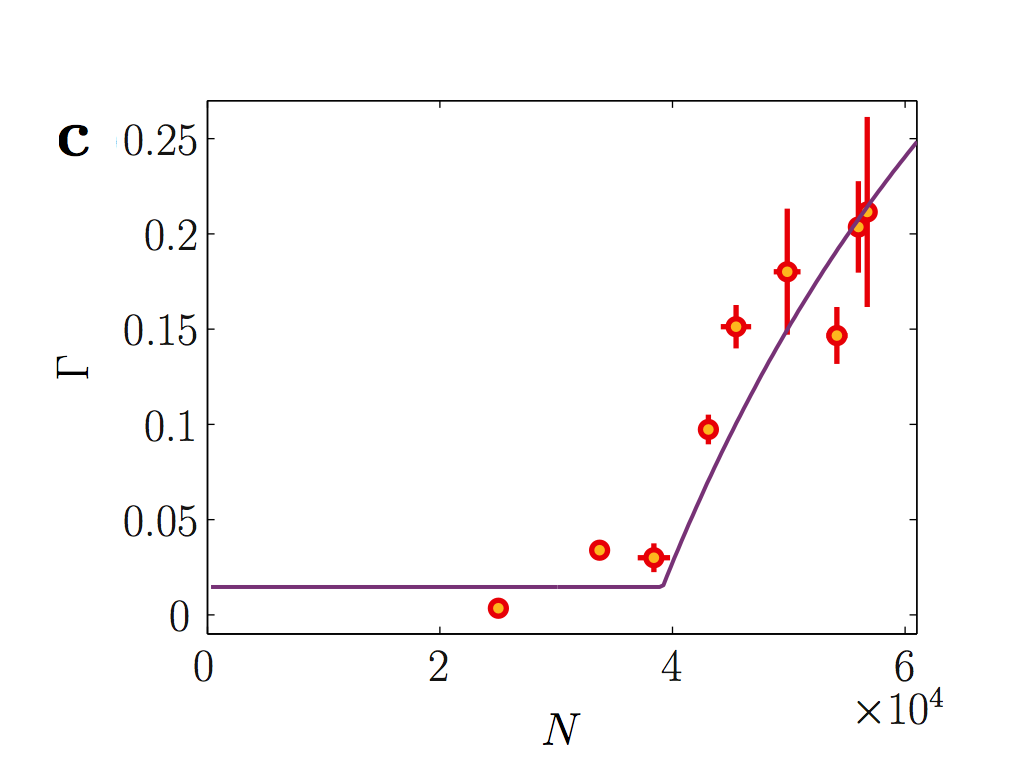}
\end{center}
\caption{ \sl \textbf{Emergence of coherence via matter--wave interference}. \textbf{(a)} Example of a density distribution after a $16$ ms in--plane expansion of two coplanar clouds initially confined in rectangular boxes of size $24\times12\,\um^2$, spaced by $d=4.5\,\um$ ($\nu_z =365$\,Hz). The region of interest considered in our analysis consists of 56 lines and 74 columns (pixel width: $0.52\,\mu$m).
\textbf{(b)} Amplitude of the 1D Fourier transform  of each line of the density distribution. Each line $y$ shows two characteristic side peaks at $\pm k_p(y)$ above the background noise, corresponding to the fringes pattern of  \textbf{a}. Here $\langle k_p \rangle =0.17(2)\,\um^{-1}$.
\textbf{(c)} Variation of the average contrast $\Gamma$ (see text for its definition) for images of gases at  $T=155$\,nK.  Error bars show the standard errors of the mean of the binned data set (with on average 3 images per point).  The solid line is a fit to the data of the function $f(N)$ defined as $f(N)= b$ for $N\leq N_{\rm c}$ and $f(N)=b + a\;(1-(N_{\rm c}/N)^{0.6})$ for $N>N_{\rm c}$.  The parameter $b$ is a constant for a data set with various $T$ taken in the same experimental conditions. Here we deduce $N_{\rm c} = 3.9\,(2)\times 10^4$, where the uncertainty range is obtained by a jackknife resampling method.}
\label{fig:fringes}
\end{figure} 
We use these interference patterns to characterize quantitatively the level of coherence of the gases initially confined in the rectangles. For each line $y$ of the pixelized image acquired on the CCD camera, we compute the $x$-Fourier transform $ \tilde{\rho}(k,y)$ of the spatial density $\rho(x,y)$ (Fig.\;\ref{fig:fringes}\,\textbf{b}). For a given $y$ this function is peaked at  a momentum $k_p(y)> 0$ that may depend (weakly) on the line index $y$. Then we consider the  function that characterizes the correlation of the complex fringe contrast $\tilde{\rho}[k_p(y),y]$ along two lines separated by a distance $d$
  \begin{equation}
\gamma(d)= \left| \; \langle \; \tilde{\rho}[k_p(y),y\,] \ \tilde{\rho}^*[k_p(y+d),y+d\,] \; \rangle\;  \right|.
 \end{equation}
Here $^*$ denotes the complex conjugation and the average is taken over the lines $y$ that overlap with the initial rectangles. If the initial clouds were two infinite, parallel lines with the same $G_1(y)$,  one would have $\gamma(d)= |G_1(d)|^2$ \cite{Polkovnikov:2006a}. Here the non-zero extension of the rectangles along $x$ and their finite initial size along $y$ make it more difficult to provide an analytic relation between $\gamma$ and the initial $G_1(r)$ of the gases. However $\gamma(d)$ remains a useful and quantitative tool to characterize the fringe pattern.   For a gas described by Boltzmann statistics, the width at $1/e$ of $G_1(r)$ is $\lT/\sqrt \pi$ and remains below $1\,\mu$m for the temperature range investigated  in this work. Since we are interested in the emergence of coherence over a scale that significantly exceeds this value, we use the following average as a diagnosis tool
\begin{equation}
\Gamma = \langle \gamma (d)\rangle, \mbox{average taken over the range } 2\,\mu{\rm m}<d<5\,\mu{\rm m} .
\label{}
\end{equation} 
For the parameter $\Gamma$ to take a value significantly different from 0, one needs a relatively large contrast on each line, and  relatively straight fringes over the relevant distances $d$, so that the  phases of the different complex contrasts do not average out.
 
For a given temperature $T$, the variation of $\Gamma$ with $N$ shows the same threshold-type behaviour as the bimodality parameter $\Delta$. One example is given in Fig.\;\ref{fig:fringes}\,\textbf{c}, from which we infer the threshold value for the atom number $N_{\rm c}(T)$ needed to observe interference fringes with a significant contrast.

\begin{figure}
\begin{center}
\includegraphics[width=14cm]{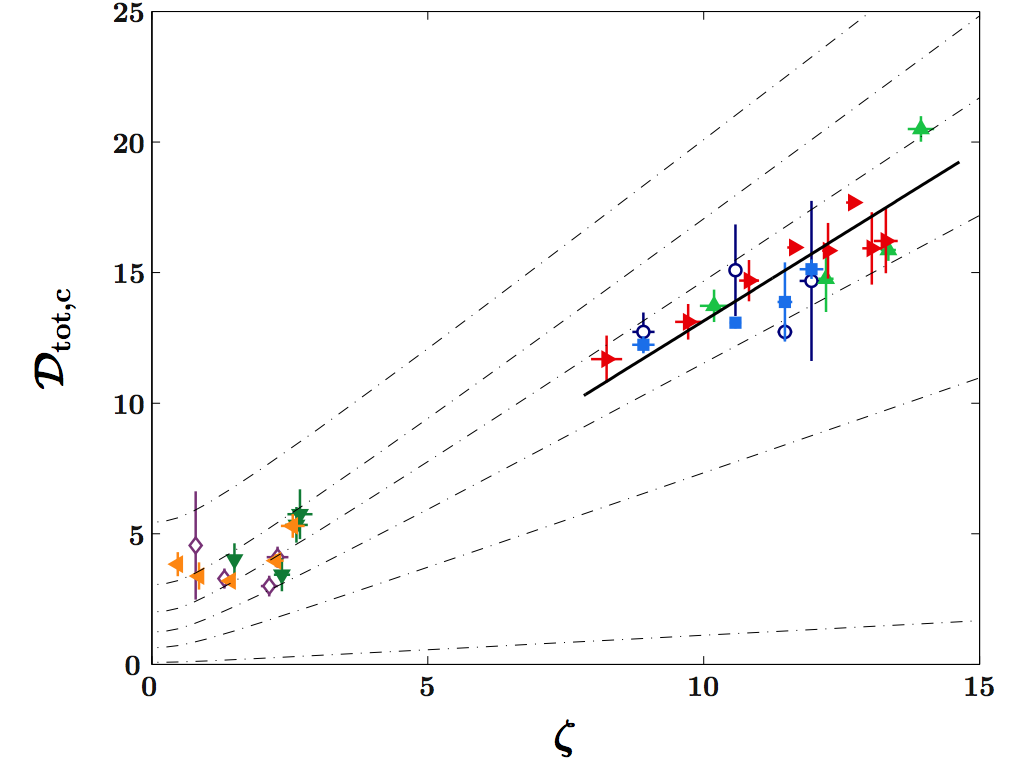} 
\end{center}
\caption{\sl \textbf{Scaling law for the emergence of coherence in a  uniform Bose gas in a quasi-2D geometry.} Variation of the threshold phase-space-density  ${\cal D}_{\rm tot,\, c}= N_{\rm c} \,\lT^2/{\cal A}$ for observing a non-Gaussian velocity distribution (full symbols) and distinct matter-wave interferences (open symbols), as a function of the dimensionless parameter $\zeta=\kB T/(h\nu_z)$. 
For velocity distribution measurements: $\nu_z=365$\,Hz: disk of radius $R=12\,\um$ (red left triangles), disk of $R=9\,\um$ (light green up triangle), square of $L=24\,\um$ (blue square), $\nu_z=1460$\,Hz: disk of $R=12\,\um$ (orange right triangles), disk of $R=9\,\um$ (dark green down triangles). For interference measurements:   $\nu_z=365$\,Hz: dark blue open circles,  $\nu_z=1460\,$Hz: violet open diamonds. 
 Error bars show the $95\%$ confidence bounds on the $N_{\rm c}$ parameter of the threshold fits to the data sets.
 The black solid line shows a linear fit to the data for $\zeta >8$, leading to ${\cal D}_{\rm tot,\, c} = 1.4\,(3)\,\zeta$. The black dash-dotted  lines show contours of identical ratios of the coherence range to the thermal wavelength $\lT$. The coherence range is evaluated by the value of $r$ at which $G_1(r) =G_1(0)/20$ (see text) and we plot (in increasing ${\cal D}_{\rm tot}$ order) ratios equal to 1, 1.2, 1.5, 2, 3 and 8. Boltzmann prediction corresponds to a ratio of $\sim 0.98$. }
 \label{fig:scaling}
\end{figure}

\vskip 5mm
\noindent{\textbf{Scaling laws for the emergence of coherence.}}
We have plotted in Fig.\;\ref{fig:scaling} the ensemble of our results for the threshold value of the total 2D phase-space-density  ${\cal D}_{\rm tot, c} \equiv N_{\rm c} \,\lT^2/{\cal A}$
 as a function of $\zeta=\kB T/h\nu_z$, determined both from the onset of bimodality as in Fig.\;\ref{fig:ToF}\,\textbf{g} (closed symbols) or from the onset of visible interference as in Fig.\;\ref{fig:fringes}\,\textbf{c} (open symbols). Two trapping configurations have been used along the $z$ direction, $\nu_z=1460\,$Hz and $\nu_z=365\,$Hz. In the first case, the $z$ direction is  nearly frozen for the temperatures studied here ($\zeta \lesssim 2$). In the second one, the $z$ direction is thermally unfrozen ($\zeta \gtrsim 8$).  All points approximately fall on a common curve, independent of the shape and the size of the gas:  
 $ {\cal D}_{\rm tot,\, c} $ varies   approximately linearly with $\zeta$ with the fitted slope 1.4\,(3) for $\zeta \gtrsim 8$\ and approaches a finite value $\sim 4$ for $\zeta \lesssim 2$. 

In the frozen case,  a majority of atoms occupy the vibrational ground state $j_z=0$ of the motion along the $z$ direction, so that ${\cal D_{\rm tot}}$ essentially represents the 2D phase-space-density associated to this single transverse quantum state.  Then for ${\cal D}_{\rm tot}\geq 1$, we know  from Eq.\;(\ref{eq:ell}) and the associated discussion that a broad component arises in $G_1$ with a characteristic length $\ell$ that increases exponentially with the phase-space-density. The observed onset of extended coherence around ${\cal D_{\rm tot}}\sim 4$ can be understood as the place where $\ell$ starts to exceed significantly $\lT$. The regime around ${\cal D}_{\rm tot}\sim 4$ is reminiscent of the presuperfluid state identified in \cite{Clade:2009,Tung:2010}. It is different from the truly superfluid  phase, which is expected  at a higher phase-space-density (${\cal D}_{\rm tot}\sim 8$) for our parameters \cite{Prokofev:2002}. Therefore the threshold $ {\cal D}_{\rm tot,\, c} $ is not associated to a true phase transition, but to a crossover where the spatial coherence of the gas increases rapidly with the control parameter $N$. 

For $\nu_z=365$\,Hz, the gas is in the ''unfrozen regime" ($\zeta \gg 1$), which could be naively thought as irrelevant for 2D physics since according to Boltzmann statistics, many vibrational states along $z$ should be significantly populated.  However thanks to the \BECp\ phenomenon presented above, a macroscopic fraction of the atoms can accumulate in the $j_z=0$ state. This happens when the total phase-space-density exceeds the threshold for \BECp\  (cf. supplementary material):
\begin{equation}
{\cal D}_{\rm tot,\,c} \approx \frac{\pi^2}{6} \,\zeta . 
\label{eq:transverseBECtext}
\end{equation}
In the limit $\zeta\to \infty$, \BECp \ corresponds to a phase transition of the same nature as the ideal gas BEC in 3D. In the  present context of our work,  we emphasize that although \BECp \ originates from the saturation of the occupation of the excited states along $z$, it also affects the coherence properties of the gas in the $xy$ plane. In particular when ${\cal D}_{\rm tot}$ rises from 0 to ${\cal D}_{\rm tot,\, c}$, the coherence length in  $xy$ increases from $\sim \lT$ (the non-degenerate result)  to $\sim a_z$, the size of the ground state of the $z$ motion. This increase can be interpreted by noting that when $\BECp$ occurs (Eq.\;\ref{eq:transverseBECtext}), the 3D spatial density in the central plane  ($z=0$) is equal to $g_{3/2}(1)/\lT^3$, where $g_s$ is the polylogarithm of order $s$ and  $g_{3/2}(1)\approx 2.612$. For an infinite uniform 3D Bose gas with this density, a true Bose-Einstein condensation occurs and the coherence length diverges. Because of the confinement along the $z$ direction, such a divergence cannot occur in the present quasi-2D case. Instead, the coherence length along $z$ is by essence limited to the  size $a_z$ of the $j_z=0$ state.  When ${\cal D}_{\rm tot}={\cal D}_{\rm tot,\, c}$ the same  limitation applies in the transverse plane, giving rise to coherence volumes that are grossly speaking isotropic. When ${\cal D}_{\rm tot}$ is increased further,  the coherence length in the $xy$ plane increases, while remaining limited to $a_z$ along the $z$ direction. 
The results shown in Fig.\;\ref{fig:scaling} are in line with this reasoning. For $\zeta \gg 1$, the emergence of coherence in the $xy$ plane occurs for a total phase-space-density  ${\cal D}_{\rm tot,\,c} \propto \zeta$, with a proportionality coefficient $\alpha=1.4\,(3)$ in good agreement with the prediction $\pi^2/6 \approx 1.6$ of Eq.\;(\ref{eq:transverseBECtext}). 

  We have also plotted in Fig.\;\ref{fig:scaling} contour lines characterizing the coherence range in terms of $\zeta$ and ${\cal D}_{\rm tot}$. Using ideal Bose gas theory, we calculated the one-body coherence function $G_1(r)$ and determined the distance $r_f$ over which it decreases by a given factor $f$ with respect to $G_1(0)$. We choose the value $f=20$ to explore the long tail that develops in $G_1$ when phase coherence emerges. The contour lines shown in Fig.\; \ref{fig:scaling} correspond to given values of $r_{20}/\lT$; they should not be considered as fits to the data, but as an indication of a coherence significantly larger than the one obtained from Boltzmann statistics (for which $r_{20}\approx \lT$). The fact that the threshold phase-space densities ${\cal D}_{\rm tot,\,c}$ follow quite accurately these contour lines \ validates the choice of tools (non Gaussian velocity distributions, matter-wave interferences) to characterize the onset of coherence. 

\begin{figure}
\begin{center}
\includegraphics[width=15cm]{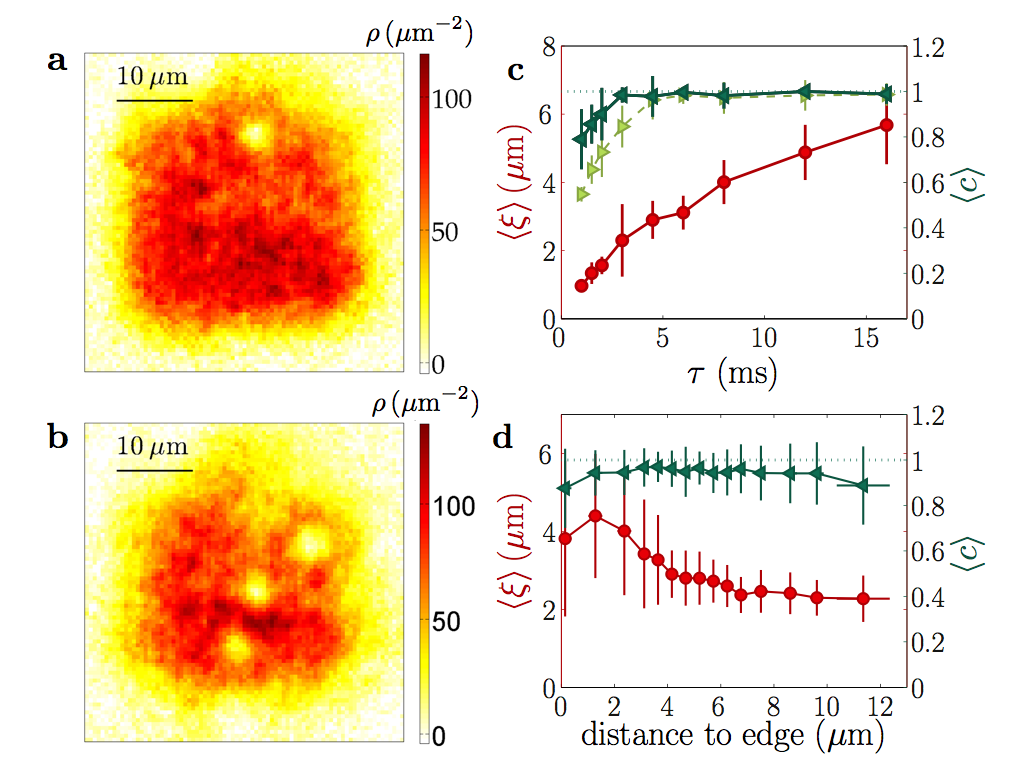} 
\end{center}
\caption{\sl \textbf{Observation of vortices}. \textbf{(a,b)} Examples of density distributions after a  3D ToF of  $\tau=4.5\,$ms for a gas initially confined in a square of size $L=30\,\um$  ($\nu_z =365\,$Hz). The two examples show respectively one ({\bf a}) and three ({\bf b})\ holes of high contrast, corresponding to topologically protected expanding vortex cores.  We fit each density hole by a hyperbolic tangent dip convoluted by a Gaussian of waist $w=1$\,\um accounting for imaging imperfections (see methods). \textbf{(c)} Evolution of the average size $\xi$ (red circles, left labels) and contrast $c$ (green triangles, right labels) of density holes with the expansion duration $\tau$.  No holes are visible for $\tau \lesssim 0.5\,$ms. Red circles and dark green left triangles are results from a fit accounting for  imaging imperfections while light green right triangles show contrast resulting from a fit without a convolution by a Gaussian.\ \textbf{(d)} Variation of the hole size $\xi$  (red circles, left labels)\ and contrast $c$ (green triangles, right labels) with the distance to the nearest edge of the box (same configuration than \textbf{a,b} : ToF  of $\tau = 4.5$\,ms for a gas in a square of $L=30\,\um$). For a distance larger than $\sim 4\,\um$, $\xi$ and $c$ are approximately independent from the vortex location.  The average values in \textbf{c} are taken over all holes independent of their positions. One point in  \textbf{c} (resp. \textbf{d}) corresponds to 15 (resp. 70) vortex fits. Error bars show standard deviations of the binned data set.}
\label{fig:vortex}
\end{figure} 

\vskip 5mm
\noindent{\textbf{Observation of topological defects.}} From now on we use the weak trap along $z$ ($\nu_z=365\,$Hz) so that the onset of extended coherence is obtained thanks to the transverse condensation phenomenon. We are interested in the regime of strongly degenerate, interacting gases, which is obtained by pushing the evaporation down to a point where the residual thermal energy $\kB T$ becomes lower than the chemical potential $\mu$ (see methods for the calculation of $\mu$ in this regime). The final box potential  is $\sim \kB \times 40$\,nK, leading to an estimated temperature of $\sim 10$\,nK, whereas the final density ($\sim50\,\mu$m$^{-2}$) leads to $\mu \approx \kB \times 14$\,nK. In these conditions, for most realizations of the experiment, defects are present in the gas. They appear as randomly located density holes after a short 3D ToF (Fig.\;\ref{fig:vortex}\,\textbf{a,b}), with a number fluctuating between 0 and 5. To identify the nature of these defects, we have performed a statistical analysis of their size and contrast, as a function of their location and of the  ToF duration  $\tau$\ (Fig \ref{fig:vortex}\,\textbf{c,d}). For a given  $\tau$, all observed holes have similar sizes and contrasts.  The core size increases  approximately linearly with $\tau$, with a nearly $100\,\%$ contrast. This favors the interpretation of these density holes as single vortices, for which the $2\pi$ phase winding around the core provides a topological protection during the ToF. This would be the case neither for vortex--antivortex pairs nor phonons, for which one would expect large fluctuations in the defect sizes and lower contrasts. 
 
\vskip 5mm
\noindent{\textbf{Dynamical origin of the topological defects.}}
In principle the vortices observed in the gas could be due to steady-state thermal fluctuations.   BKT theory indeed predicts that vortices should be present in an interacting 2D Bose gas around the superfluid transition point \cite{Kosterlitz:1973}. Such ``thermal'' vortices have been observed in non-homogeneous atomic gases, either interferometrically \cite{Hadzibabic:2006} or as density holes in the trap region corresponding to the critical region \cite{Choi:2013b}. However, for the large and uniform phase-space densities that we obtain at the end of the cooling process  ($\rho \lT^2 \geq 100$), Ref. \cite{Giorgetti:2007} predicts a  vanishingly small probability of occurrence for such thermal excitations. This  supports a dynamical origin for the observed defects. 

To investigate further this interpretation, we can vary the two times that characterize the evolution of the gas, the duration of evaporation $\tev$  and the hold duration after evaporation $\tho$ (see Fig.\;\ref{fig:setup}\,\textbf{a}). For the results presented in this section, we  fixed $\tho=500$\,ms and studied the evolution of the average vortex number $\Nv$ as a function of $\tev$. The corresponding data, given in Fig.\;\ref{fig:kz}\,\textbf{a}, show a decrease of $\Nv$ with $\tev$, passing from $\Nv \approx 1$ for $\tev=50$\,ms  to $\Nv \approx 0.3$ for $\tev=250$\,ms. For longer evaporation times, $\Nv$ remains approximately constant around $0.35\,(5)$.

The decrease of $\Nv$ with $\tev$ suggests that the observed vortices are nucleated via a Kibble--Zurek (KZ) type mechanism \cite{Kibble:1976, Zurek:1985, Anglin:1999}, occurring when the transition to the phase coherent regime is crossed. However applying  the KZ formalism to our setup is not straightforward. In a weakly interacting, homogeneous 3D Bose gas, BEC occurs when the 3D phase-space-density  reaches the critical value $g_{3/2}(1)$. For our quasi-2D geometry, transverse condensation occurs when the 3D phase-space-density in the central plane $z=0$ reaches this value. At the transition point, the KZ formalism relates the size of phase-coherent domains to the cooling speed $\dot T$.  For fast cooling, KZ theory predicts domain sizes for a 3D fluid that are smaller than or comparable to the thickness $a_z$ of the lowest vibrational state along $z$;  it can thus provide a good description of our system. For a slower cooling, coherent domains much larger than $a_z$ would be expected in 3D at the transition point. The 2D nature of our gas leads in this case to a reduction of the in-plane correlation length.  In the slow cooling regime, we thus expect to find an excess of topological defects with respect to the KZ prediction  for standard 3D BEC.

More explicitly we expect for  fast cooling, hence short $\tev$, a power-law decay $\Nv \propto \tev^{-d}$ with an exponent $d$ given by the KZ formalism for 3D BEC. The fit of this function to the measured variation of $\Nv$ for $\tev \le 250\,$ms leads to $d=0.69\,(17)$ (see Fig.\,\ref{fig:kz}\,\textbf{a}). This is in good agreement with the prediction $d=2/3$ obtained from the critical exponents of the so-called "F model"  \cite{Hohenberg:1977}, which is believed to describe the universality class of the 3D BEC phenomenon. For comparison, the prediction for a pure mean-field transition, $d=1/2$, is notably lower than our result.

For longer $\tev$, the above described excess of vortices due to the quasi-2D geometry should translate in a weakening of the decrease of $\Nv$ with $\tev$. The non-zero plateau observed in Fig.\,\ref{fig:kz}\,\textbf{a} for $\tev \ge 250\,$ms may be the signature of such a  weakening. Other mechanisms could also play a role in the nucleation of vortices for slow cooling. For example due to the box potential residual rugosity, the gas could condense into several independent patches of fixed geometry, which would merge later during the evaporation ramp and stochastically form vortices with a constant probability.

 \begin{figure}
\begin{center}
\includegraphics[trim=0 210 0 210,width=15cm]{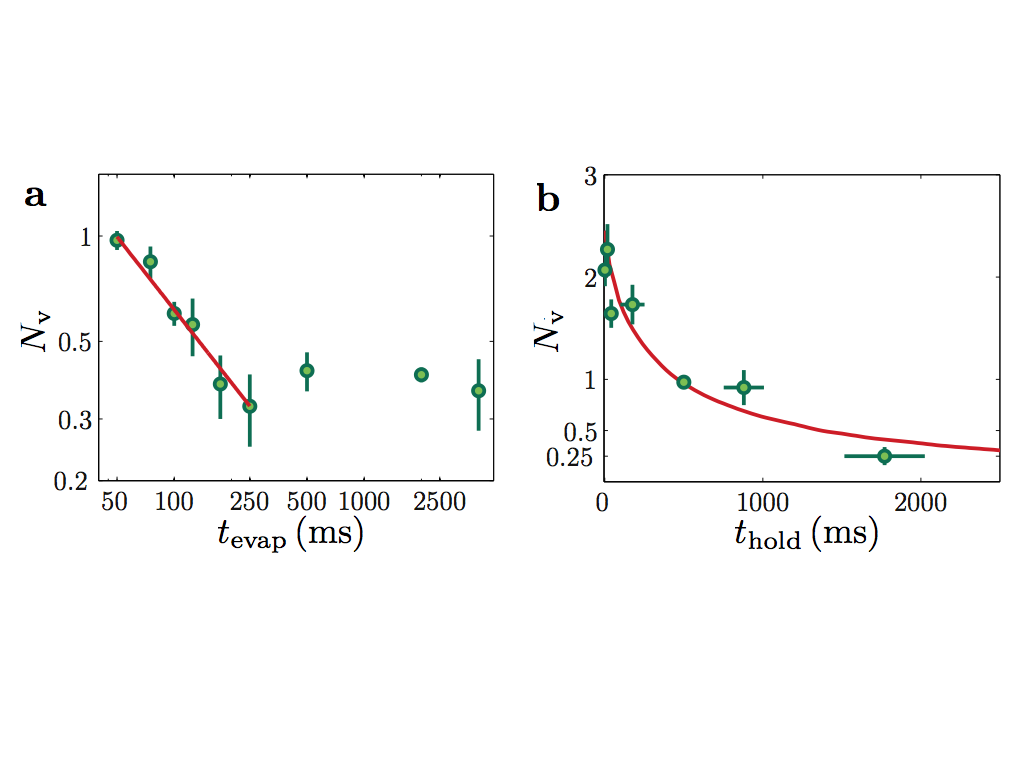} 
\end{center}
\caption{\sl \textbf{Dynamics of vortex nucleation.} 
\textbf{(a)} Circle symbols: evolution of the mean vortex number $\Nv$ with the quench time $\tev$ (fixed $\tho=500\,$ms)  for a gas initially confined in a square of size $L=30\,\um$  ($\nu_z =365\,$Hz) and observed after a 3D ToF of $\tau = 4.5\,$ms.  The number of images per point ranges from 37 to 233, with a mean of 90. We restrict to $t_{\rm evap}\geq 50$\,ms to ensure that local thermal equilibrium is reached at any time during the evaporation ramp \cite{Ketterle:1996}. Red line: fit of a power-law decay to the short time data ($t_{\rm evap}\leq 250\,$ms), giving the exponent $d=0.69\,(17)$.   The uncertainty range on $d$ is the $95\%$ confidence bounds of a linear fit to the evolution of $\log(\Nv)$ with  $\log(\tev)$. For longer quench times, the mean vortex numbers are compatible with a plateau at $N_{\rm v}=0.35\,(5)$. \textbf{(b)} Circle symbols: evolution of the mean vortex number $\Nv$ with the hold time $\tho$ (fixed $\tev=50\,$ms) in the same experimental configuration as \textbf{a}. The number of images per point ranges from 24 to 181, with a mean of 59. In both figures error bars are obtained from a bootstrapping approach. Red line: results from a model describing the evolution of an initial number of vortices $N_{\rm v,0}=2.5(2)$ in the presence of a phenomenological damping coefficient \cite{Fedichev:1999}. The inferred superfluid fraction is $0.94\,(2)$.  Confidence ranges  on these parameters are obtained from a $\chi^2$-analysis.  
   }
\label{fig:kz}
\end{figure}

\vskip 5mm
\noindent{\textbf{Lifetime of the topological defects.}} The variation of the number of vortices $\Nv$ with the hold time $\tho$ allows one to study the fate of vortices that have been nucleated during the evaporation. We show in Fig.\;\ref{fig:kz}\,\textbf{b} the results obtained when fixing the evaporation to a short value $\tev=50$\,ms. We observe a decay of $\Nv$ with the hold time, from $\Nv=2.3$ initially to $0.3$ at long $\tho$ (2\,s). To interpret this decay, we modeled the dynamics of the vortices in the gas with two ingredients: (i) the conservative motion of a vortex in the velocity field created by the other vortices, including the vortex images from the boundaries of the box potential \cite{Donnelly:1991}, (ii) the dissipation induced by the scattering of thermal excitations by the vortices, which we describe phenomenologically by a friction force that is proportional to the non-superfluid fraction of atoms in the gas    
\cite{Fedichev:1999}.  During this motion, a vortex annihilates when it reaches the edge of the trap or encounters another vortex of opposite charge. The numerical solution of this model leads to a non-exponential decay of the average number of vortices,  with details that depend on the initial number of vortices and their locations.

Assuming a uniform random distribution of vortices at the end of the evaporation, we have compared the predictions of this model to our data. It gives the following values of the two adjustable parameters of the model, the initial number of vortices $N_{\rm v,0}=2.5\,(2)$ and the superfluid fraction  $0.94\,(2)$; the corresponding prediction is plotted as a continuous line in Fig.\;\ref{fig:kz}\,\textbf{b}.  We note that at short $\tho$, the images of the clouds are quite fuzzy, probably because of  non-thermal phononic excitations produced (in addition to vortices) by the evaporation ramp. The   difficulty to precisely count vortices in this case leads to fluctuations of $\Nv$ at short $\tho$ as visible in Fig.\;\ref{fig:kz}\,\textbf{b}. The choice $\tho=500$\,ms  in Fig.\;\ref{fig:kz}\,\textbf{a} was made accordingly.

The finite lifetime of the vortices in our sample points to a general issue that one faces in the experimental studies on the KZ mechanism. In principle the KZ formalism gives a prediction on the state of the system just after crossing the critical point. Experimentally we observe the system at a later stage, at a moment when the various domains have  merged, and we detect the topological defects  formed from this merging. In spite of their robustness, the number of vortices is not strictly conserved after the crossing of the transition and its decrease depends on their initial positions. A precise comparison between our results and KZ theory should take this evolution into account, for example using stochastic mean-field methods  \cite{Bisset:2009b,Mathey:2010,Das:2012,Cockburn:2012}.

\section*{Discussion} 

Using a box-like potential created by light, we developed a setup that allowed us to investigate the quantum properties of atomic gases in a uniform quasi-2D configuration. Thanks to the precise control of atom number and  temperature, we characterized the regime for which phase coherence emerges in the fluid.  The uniform character of the gas allowed us to disentangle the effects of ideal gas statistics for in-plane motion, the notion of transverse condensation along the strongly confined direction, and the role of interactions. This is to be contrasted with previous studies that were performed in the presence of a harmonic confinement in the plane, where these different phenomena could be simultaneously present in the non-homogeneous atomic cloud. 

For the case of a weakly interacting gas considered here, our observations highlight the importance of Bose statistics in the emergence of extended phase coherence. This  coherence is already significant for phase-space densities  ${\cal D}_0\sim 3-4$, well below the values required for (i) the superfluid BKT transition and (ii) the full Bose--Einstein condensation in the ground state of the box. For our parameters, the latter transitions are expected around the same phase-space-density ($\sim 8-10$) meaning that when the superfluid criterion is met, the coherence length set by Bose statistics is comparable to the box size.  

By cooling the gas further, we entered the  regime where interactions dominate over thermal fluctuations. This allowed us to visualize with a very good contrast the topological defects (vortices) that are created during the formation of the macroscopic matter-wave, as a result of a Kibble--Zurek type mechanism. Here we focused on the relation between the vortex number and the cooling rate. Further investigations could include correlation studies on vortex positions, which can shed light on their nucleation process and their subsequent evolution \cite{Freilich:2010}. 

Our work motivates future research in the direction of strongly interacting 2D gases \cite{Ha:2013}, for which the order of the various transitions could be interchanged. In particular the critical ${\cal D}$ for the BKT transition should decrease, and reach ultimately the universal value of the "superfluid jump",   ${\cal D}=4$ \cite{Nelson:1977}. In this case, the emergence of extended coherence in the 2D gas would be essentially driven by the interactions. Indeed once the superfluid transition is crossed,  the one-body correlation function is expected to decay very slowly,  $G_1(r) \propto r^{-\alpha}$, with $\alpha< 1/4$. It would be  interesting to revisit the statistics of formation of quench-induced topological defects in this case, for which significant deviations to the KZ power-law scaling have been predicted \cite{Jelic:2011,Dziarmaga:2013}.

\section*{Methods}
\noindent{\textbf{Characterization of the box--like potential.}}
We create the box-like potential in the $xy$ plane using a  laser beam that is blue-detuned with respect to the \Rb resonance. At the position of the atomic sample, we image a dark mask placed on the path of the laser beam. This mask is realized by a metallic deposit on a wedged, anti-reflective coated glass plate. We characterize the box-like character of the resulting  trap in two ways. (i) The flatness of the domain where the atoms are confined is characterized by the root mean square intensity fluctuations of the inner dark region of the beam profile.  The resulting variations of the dipolar potential are $\delta U/U_{\rm box} \sim 3\%$, where $U_{\rm box}$ is the potential height on the edges of the box.  The ratio $\delta U/\kB T$  varies from $\sim 40 \%$ at the loading temperature to $\sim 10\%$ at the end of the evaporative cooling (ramp of $U_{\rm box}$, see below). In particular, it is of $\sim 20\%$ at the transverse condensation point for the configuration in which the vortex data have been acquired.
(ii)  The sharp spatial variation  of the potential at the edges of the box-like trapping region is characterized by the exponent $\alpha$ of a power-law fit $U(r) \propto r^\alpha$ along a radial cut. We restrict the fitting domain to the central region where $U(r)<U_{\rm box}/4$ and find $\alpha \sim 10$--$15$, depending on the size and the shape of the box.

\vskip 5mm
\noindent{\textbf{Imaging of the atomic density distribution.}}
We measure the atomic density distribution in the $xy$ plane using resonant absorption imaging along $z$. We use two complementary values for the probe beam intensity $I$. First we use a conventional low intensity technique with $I/I_{\rm sat} \approx 0.7$, where $I_{\rm sat}$ is the  saturation intensity of the Rb resonance line,  with a probe pulse duration of $20\,\mu$s. This procedure enables a reliable detection of low density atomic clouds, but it is unfaithful for high density ones, especially in the 2D geometry due to multiple scattering effects between neighboring atoms \cite{Chomaz:2012}. We thus complement it by a high intensity technique inspired from \cite{Reinaudi:2007}, in which we  apply a short pulse of $4\,\mu$s of an intense probe beam  with  $I/I_{\rm sat} \approx 40$. ToF bimodality measurements (where the cloud is essentially 3D at the moment of detection) were performed with the low intensity procedure. This was also the case for the matter-wave interference measurements, for which we reached a better fringe visibility in this case. In situ images in Fig.\,\ref{fig:setup} and all data related to vortices (\emph{e.g.} Fig.\;\,\ref{fig:vortex}\,\textbf{a},\textbf{b}) in the strongly degenerate gases were taken with high intensity imaging. We estimate the uncertainty on the atom number to be of $20\%$.

\vskip 5mm
\noindent{\textbf{Ideal gas description of trapped atomic samples.}}
We consider a gas of $N$ non-interacting bosonic particles confined in a square box of size $L$ in the $xy$ plane, and in a harmonic potential well of frequency $\nu_z$ along $z$. The eigenstates of the single-particle Hamiltonian are labelled by three integers $j_x,j_y \geq 1$, $j_z \geq 0$: 
\begin{equation}
\psi_\vj(\bs r)=\frac{1}{L\sqrt{a_{z}}}\sin(\pi j_x x/L)\;\sin(\pi j_y y/L) \;\chi_{j_z}(z/a_z),
\label{}
\end{equation}
where $a_z=(h/m\nu_z)^{1/2}/(2\pi)$ and $\chi_j$ is the normalized $j$-th Hermite function. Their energies and occupation factors are 
\begin{eqnarray}
\label{eq:Ej}
E_{\vj} &=&\frac{h^2}{8m L^2} (j_x^2 + j_y^2-2) + j_z \,h \nu_{z},\\
\label{eq:popbose}
n_{\vj} &=&\frac{1}{\exp\left[(E_{\vj}-\mu)/\kB T\right]-1},
\end{eqnarray} 
where $\mu < 0$ is the chemical potential of the gas and $N=\sum_\vj n_{\vj}$. 
The average value of any one-body observable $\hat{A}$ can then be calculated:
\begin{equation} 
\label{eq:Amean}
\langle \hat{A} \rangle =\frac{1}{N} \sum_{\vj} n_{\vj} \;\langle \psi_\vj | \hat{A} |\psi_\vj \rangle.
\end{equation} 

\vskip 5mm
\noindent{\textbf{Estimation of the interaction energy for weakly interacting gases.}}
We estimate the local value of the interaction energy per particle $\epsilon_{\rm int}=(2\pi\hbar^2 a/m) \rho^{\rm (3D)}(\bs r)$, where $a=5.1$\,nm is the 3D scattering length  characterizing $s$-wave interactions for $^{87}$Rb atoms and   $\rho^{\rm (3D)}(\bs r)$ the spatial 3D density estimated using the ideal gas description. It is maximal at trap center $\bs r =0$. 
For example, using a typical experimental condition with $N=40\,000$ atoms in a square box of size $L=24\,\mu$m at $T=200$\,nK, we find a maximal 3D density of $\rho^{(3D)}(0)=13.8\,\mu$m$^{-3}$. The mean-field interaction energy for an atom localized at the center of cloud is then $\epsilon_{\rm int}=\kB \times 2.1\,$nK. We note that $\epsilon_{\rm int}$ is negligible compared to $\kB T$ and $h\nu_z$ for all atomic configurations corresponding to the onset of an extended phase coherence. In this case the interactions play a negligible role in the 2D ToF expansion that we use to reveal matter-wave interferences.

\vskip 5mm
\noindent{\textbf{Temperature calibration.}}
All temperatures indicated in the paper are deduced from the value of the box potential, assuming that the evaporation barrier provided by $U_{\rm{box}}$ sets the thermal equilibrium state of the gas. This  
hypothesis was tested, and the relation between $T$ and $U_{\rm{box}}$ calibrated, using atomic assemblies with a negligible interaction energy. For these  assemblies, we compared the variance of their velocity distribution $\Delta v^2$ obtained from a ToF measurement to the prediction of Eq.\;(\ref{eq:Amean}). The calibration obtained from this set of measurements can be empirically written as
\begin{equation}
\label{eq:TU} 
T(U_{\rm{box}})=T_0 \left(1-{\rm e}^{- U_{\rm{box}}/ (\eta \,\kB T_0)}\right),
\end{equation}
where the values of the dimensionless parameter  $\eta$ and of the reference temperature $T_0$ slightly depend on the precise shape of the trap. For the square trap of side $24\,\mu$m we obtain $T_0=191\,(6)$\,nK and $\eta=3.5\,(3)$. The reason for which $T$ saturates when the box potential increases to infinity is due to the residual evaporation along the vertical direction, above the barrier created by the horizontal Hermite--Gauss beam.

\vskip 5mm
\noindent{\textbf{Power exponent for fitting $N_c$.}}
We estimate the behavior of $\Delta (N)$ at fixed $T$ using Bose law for a ideal gas. We compute from (\ref{eq:Amean}) the equilibrium velocity distribution $\tilde{\rho}({\bs v}). $ Then we estimate the spatial density after a ToF of duration $\tau$ (for a disk trap of radius $R$) via 
$\rho({\bs r}) \propto \tilde{\rho}({\bs r}/\tau) * \Theta{\left(r \leqslant R\right)}$
where $*$ stands for the convolution operator and $\Theta$ for the Heaviside function. We fit $\rho({\bs r})$ to a double Gaussian and compute the atom fraction in the sharpest Gaussian $\Delta$, similarly to the processing of experimental data.  To simulate our experimental results, we consider $\nu_z=350\,$Hz, $R=12\,\um$,  $\tau =14\,$ms and $T$ varying from $100$ to $250\,$nK. For a given $T$, we record $\Delta$ while varying the total atom number $N$ from 0.06 to 4 times the theoretical critical number for $\BECp$ $N_{\rm c, th} = \zeta (\pi^2/6) {\cal A}/\lT^2$ (see Sup. Mat.). We fit $\Delta(N)$ between $N_{\rm min}=0.06 \;N_{\rm c, th}$ and a varying $N_{\rm max}$ in $1.1$ -- $4 \;N_{\rm c, th}$, to 
$f(N) =  \left( 1 - \left(N_c/N \right)^\alpha\right)$
with $N_c$ as a free parameter  and a fixed $\alpha$. 
For all considered $T$ and $N_{\rm max}$, choosing $\alpha=0.6$ provides both a good estimate of $N_c$ (between $0.93$ and $0.99\;N_{\rm c, th}$) and a satisfactory fit  (average coefficient of determination $0.94$).

\vskip 5mm
\noindent{\textbf{Chemical potential in the degenerate interacting regime.}}
To compute the chemical potential $\mu$ of highly degenerate interacting gases, we perform a $T=0$ mean-field analysis. We solve numerically the 3D Gross--Pitaevskii equation in imaginary time using a split--step method, and we obtain the macroscopic ground state wave-function $\psi(\bs r)$.
Then we calculate the different energy contributions at $T=0$ -- namely the potential energy $E_{\rm pot}$, the kinetic energy $E_{\rm kin}$ and interaction energy $E_{\rm int}$ --  by integrating over space :
\begin{eqnarray}
E_{\rm pot}
& =& \frac{N}{2}m\omega_z^{2}\, \int  z^{2}|\psi(\bs r)|^2 \mathrm{d}^{3}r, \\
E_{\rm kin}
&= &\frac{N\,\hbar^2 }{2 m} \int \left|\nabla\psi(\bs r)\right|^2 \mathrm{d}^{3}r  ,\\
E_{\rm int}  &=& N^2\,\frac{2\pi \hbar^2 a}{m}\,  \int \left|\psi(\bs r)\right|^4 \mathrm{d}^{3}r ,
\end{eqnarray} 
with $\omega_z=2\pi \nu_z$ and $\hbar =h/(2\pi)$.   
We obtain the value of the chemical potential $\mu$ by taking the derivative of the total energy with respect to $N$ and subtracting the single-particle ground state energy:
\begin{equation}
\mu = \frac{1}{N}\left(E_{\rm pot}+E_{\rm kin} +2E_{\rm int}\right) -\frac{h^2}{4mL^2} -  \frac{1}{2}h \nu_z.
\end{equation}
In the numerical calculation, we typically use time steps of $10^{-4}\,$ms and compute the evolution for $10\,$ms. The 3D grid contains $152\times152\times32$ voxels, with a voxel size $0.52\times 0.52\times 0.26\,\um^3$.

\vskip 5mm
\noindent{\textbf{Analysis of the density holes created by the vortices.}} We first calculate the normalized density profile $\rho/\bar \rho$ where the average $\bar \rho$ is taken over the set of images with the same ToF duration $\tau$. Then we look for density minima with a significant contrast and size. Finally for each significant density hole, we select a square region centered on it with a size that is $\sim 3$ times larger than the average hole size for this $\tau$. In this region, we fit the function 
\begin{equation}
\label{eq:rho_0}
A_0\left[1-c+c\tanh\left(\sqrt{x^2+y^2}/\xi\right)\right]
\end{equation}
to the normalized density profile, where $A_0$ accounts for density fluctuations. We also correct for imaging imperfections (finite imaging resolution and finite depth of field) by performing a convolution of the function defined in Eq.(\ref{eq:rho_0}) by a Gaussian of width $1\,\mu$m, which we determined from a preliminary analysis.

\vskip 5mm
\noindent{\textbf{Acknowledgments.}
We thank J. Palomo and D. Perconte for the realization of the intensity masks and  Zoran Hadzibabic for several useful discussions.
This work is supported by IFRAF, ANR (ANR-12- 247 BLANAGAFON), ERC (Synergy grant UQUAM) and the Excellence Cluster CUI. L. Ch.  and L. Co. acknowledge the support from DGA, and C. W.  acknowledges the support from the EU (PIEF-GA-2011- 299731).

\newpage

\begin{center}
\textbf{\LARGE Supplementary material: \\
\vskip 5mm
Transverse condensation and 2D coherence}
\end{center}

\vskip 10mm
Most of the experimental data have been taken with a confinement frequency along the $z$ axis $\nu_z=365\,$Hz $ = (\kB/h)\, 18\,$nK. These data thus lie in the regime $\zeta = \kB T/h\nu_z > 1$. However, thanks to Bose statistics, one can still reach a situation where the surface density $\rho^{\rm 2D}_0$ associated to the ground state of the $z$ motion $|j_z=0\rangle$  is comparable to the total surface density $\rho^{\rm 2D}_{\rm tot}$. Indeed  the surface density $\rho^{\rm 2D}_{\rm exc}=\rho^{\rm 2D}_{\rm tot}-\rho^{\rm 2D}_0$ that can be accumulated in the excited states of the $z$ motion is bounded \cite{vand97}. When $\rho^{\rm 2D}_{\rm tot}$ largely exceeds this bound, essentially every additional atom accumulates in the ground state of the $z$ motion $|j_z=0\rangle$. In this supplementary material, we analyse this transverse condensation  phenomenon (\BECp) and show that the coherence length in the $xy$ plane is also significantly affected when \BECp\ occurs.

\section{Thermal equilibrium of an ideal Bose gas in a quasi-2D  geometry}

We consider a gas of $N$ non-interacting bosonic particles confined in a square box of size $L$ in the $xy$ plane and in a harmonic potential well of frequency $\nu_z$ along $z$. We use Dirichlet boundary conditions in the $xy$ plane so that the eigenstates of the single-particle Hamiltonian are labeled by two quantum numbers $j_x,j_y$ describing the state in the $xy$ plane:
\begin{equation}
\frac{1}{L}\,\sin(\pi j_x x/L)\,\sin(\pi j_y y /L), \qquad j_x,j_y \mbox{ strictly positive integers}, 
\label{}
\end{equation}
and the quantum number $j_z \geq 0$ describing the vibrational state along $z$. Putting by convention the energy of the ground state at zero, the energies and occupation factors of these energy levels are
\begin{eqnarray}
\label{eq:popbose}
\label{eq:Ej}
E_{\bs j} &=&(j_x^2+j_y^2-2)\frac{h^2 }{8mL^2} + j_z \,h \nu_{z}, \label{eq:quantum_states}\\
n_{\bs j} &=&\frac{1}{\exp\left[(E_{\bs j}-\mu)/\kB T\right]-1},
\end{eqnarray} 
where $\bs j=(j_x,j_y,j_z)$, $\hbar=h/(2\pi)$, $\mu < 0$ is the chemical potential of the gas and the total atom number in the gas is
\begin{equation}
N=\sum_{\bs j} n_{\bs j}.
\label{}
\end{equation} 

The number of atoms in the $j_z$ vibrational state is 
\begin{equation}
N_{j_z}=\sum_{j_x,j_y} n_{\bs j}
\label{}
\end{equation}
so that the 2D phase-space-density associated to that state is 
\begin{equation}
{\cal D}_{j_z}^{\rm 2D}= \frac{\lT^2}{L^2} \sum_{j_x,j_y} n_{\bs j} .
\label{}
\end{equation}
Turning the discrete sum over $j_x,j_y$ into an integral in the limit where $L/\lT$ is very large, we find
\begin{equation}
{\cal D}_{j_z}^{\rm 2D}= \frac{\lT^2}{4\pi^2} \int n_{\bs j}\;\D^2 k= - \ln \left(1-Z\E^{-j_z/\zeta}  \right), 
\label{}
\end{equation}
where $Z=\mu/\kB T$ and $\zeta=\kB T/ h\nu_z$. The total 2D phase-space-density is
\begin{equation}
{\cal D}_{\rm tot}^{\rm 2D}=\frac{N \lT^2}{L^2}=\sum_{j_z} {\cal D}_{j_z}^{\rm 2D}.
\label{eq:PSDtot}
\end{equation}

\section{Transverse condensation in a quasi-2D geometry}

In order to investigate the transverse condensation phenomenon, we use a treatment very similar to that of  usual 3D Bose-Einstein condensation. We  focus on the case $\zeta \gg 1$ where many vibrational states along the $z$ direction are populated. Using the semi-classical approximation that consists in replacing the discrete sum in (\ref{eq:PSDtot}) by an integral over $j_z$, we calculate the total  phase-space-density :
\begin{eqnarray}
{\cal D}_{\rm tot}^{\rm 2D} &=& -\sum_{j_z=0}^{+\infty} \ln\left(1-Z\E^{-j_z/\zeta}  \right) 
\label{eq:discrete}\\
 &\approx& -\int_0^{+\infty} \ln\left(1-Z\E^{-u/\zeta}  \right)\;\D u \label{eq:integral} \\
 & = & \sum_{n=1}^{+\infty }\int_0^{+\infty} \frac{Z^n}{n}\E^{-nu/\zeta}  \;\D u 
 = \zeta \sum_{n=1}^{+\infty } \frac{Z^n}{n^2} 
  =  \zeta\, g_2(Z)  . 
\label{eq:transverseBEC}
\end{eqnarray}
Since $Z <1$ and $g_2(Z)$ remains finite when $Z\to 1$ [$g_2(1) = \pi^2/6$], this semi-classical approximation leads to the paradoxical result that for a given $\zeta$, the total 2D phase-space-density is bounded from above by $\zeta \pi^2/6$. 

\begin{figure}[p]
\begin{center}
\includegraphics{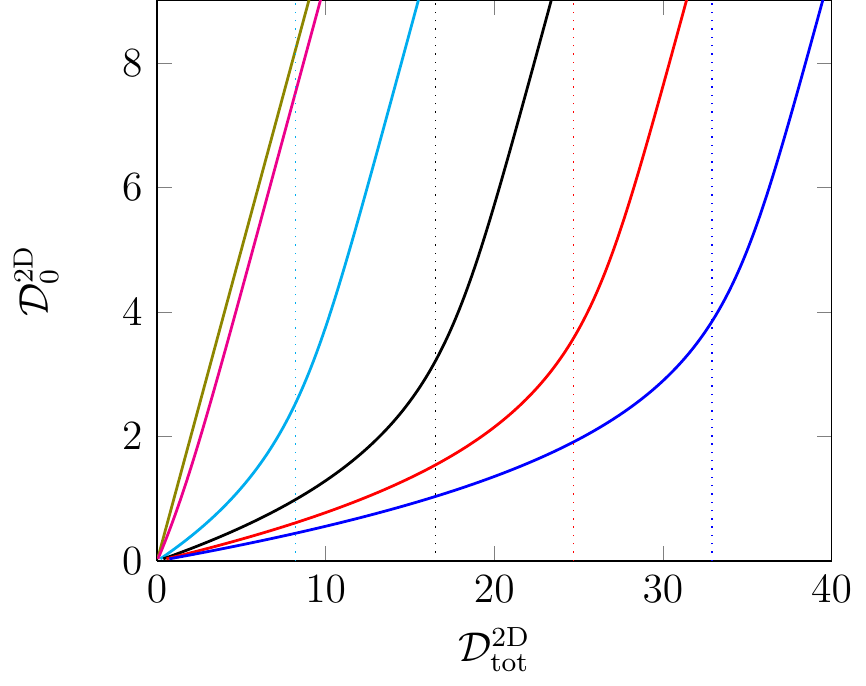}
\includegraphics{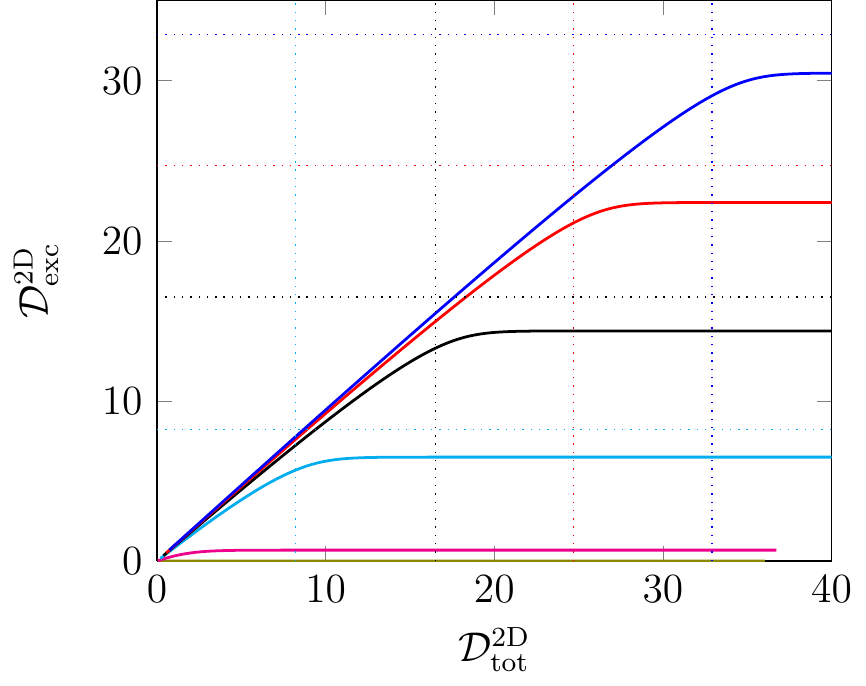}
\end{center}
\caption{\sl Top: Variations of the 2D phase-space-density in the $|j_z=0\rangle$ ground state of the $z$ motion with the total 2D phase space-density ${\cal D}_{\rm tot}^{\rm 2D}$. Bottom: Variations of the phase-space-density of the excited states of the $z$ motion with ${\cal D}_{\rm tot}^{\rm 2D}$. The calculation is made for an ideal gas confined in a square box in the limit $L/\lT \to \infty$. The values of $\zeta$ are: 0.1 (olive), 1 (magenta), 5 (cyan), 10 (black), 15 (red), 20 (blue). The dotted lines indicate the critical phase-space-density $\zeta\,\pi^2/6$ (same color code).}
\label{fig:2D_PSDs}
\end{figure}

The paradox is lifted by noticing that when the fugacity $Z$ approaches 1, the population of the lowest vibrational state $|j_z=0\rangle$ is not properly accounted for when one replace the discrete sum in (\ref{eq:discrete}) by the integral (\ref{eq:integral}). More precisely within this semi-classical approximation, when the total phase-space-density approaches the value $\zeta \pi^2/6$, the result above must be replaced by
\begin{equation}
{\cal D}_{\rm tot}^{\rm 2D} = {\cal D}^{\rm 2D}_0 + {\cal D}^{\rm 2D}_{\rm exc}
\label{}
\end{equation}
with
\begin{equation}
\mbox{for }Z \approx 1:\qquad {\cal D}^{\rm 2D}_0 =-\ln(1-Z), \qquad {\cal D}^{\rm 2D}_{\rm exc}= \zeta\, \pi^2/6.
\label{}
\end{equation}
Therefore, when the total phase-space-density ${\cal D}_{\rm tot}^{\rm 2D}$ is significantly larger than $\zeta\, \pi^2/6$,  the phase-space-density ${\cal D}^{\rm 2D}_{\rm exc}$ saturates and additional atoms accumulate essentially in the $|j_z=0\rangle$ state  (see Fig.\;\ref{fig:2D_PSDs}) \cite{vand97}. 

Let us estimate the 2D phase-space-density associated to the ground state of the $z$ motion when ${\cal D}_{\rm tot}^{2D}$ reaches the threshold for transverse condensation 
\begin{equation}
{\cal D}_{\rm tot,c}^{2D}= \zeta\;\pi^2/6.
\label{eq:BECp}
\end{equation}
 At this point, the population of the state  $|j_z=0\rangle$ is significantly different from that of $|j_z=1\rangle$, so that it cannot be accounted for properly by the integral (\ref{eq:integral}). This implies that the chemical potential is on the order of $-h \nu_z$, i.e., $Z\sim \E^{-1/\zeta}$ and we thus predict
\begin{equation}
{\cal D}^{2D}_0=-\ln(1-Z)\sim\ln(\zeta)
\label{eq:psd0_c}
\end{equation}
More precisely, a numerical calculation (see Table 1) gives at the condensation point
\begin{equation}                      
{\cal D}^{2D}_0\approx \ln (\zeta)+0.9.
\label{}
\end{equation} 
for $\zeta$ between 5 and 20.

\begin{table}[b]
\begin{center}
\begin{tabular}{|r|c|c|c|c|c|c|}
$\zeta$ & ${\cal D}^{\rm 2D}_{\rm tot,c}$ & $Z$ & ${\cal D}_{0}^{\rm 2D}$ & $r_{20}/\lambda_T$ & $l_{\rm coh}/\lambda_T$ &$l_{\rm coh}/a_{\rm ho}$\\
\hline  
5 & 8.2 & 0.921 &  2.5 & 2.2 & 0.80 & 0.90 \\
10 & 16.5 & 0.959 & 3.2 & 2.6 & 1.07 & 0.85 \\
15 & 24.7 & 0.972 & 3.6 & 2.9 & 1.27 & 0.82 \\
20 & 32.9 & 0.979 & 3.8 & 3.1 & 1.43 & 0.80
\end{tabular}
\end{center}
\label{tab:typical}
\caption{\sl Values of relevant parameters at the point where $\BECp$ occurs.}
\end{table}

In the limit $\zeta \to \infty$, this transverse condensation phenomenon (\BECp) constitutes a phase transition (see Fig.\;\ref{fig:phase_transition}): denoting the transversely condensed fraction as $f_0={\cal D}^{\rm 2D}_0/{\cal D}^{\rm 2D}_{\rm tot}$ and the reduced total phase-space-density as $x={\cal D}^{\rm 2D}_{\rm tot}/{\cal D}^{\rm 2D}_{\rm tot,\;c}$, we find in this limit:
\begin{equation}
f_0=0 \mbox{ if }x<1, \qquad f_0=1-\frac{1}{x}\mbox{ if }x\geq 1.
\label{}
\end{equation}

\begin{figure}[t]
\begin{center}
\includegraphics{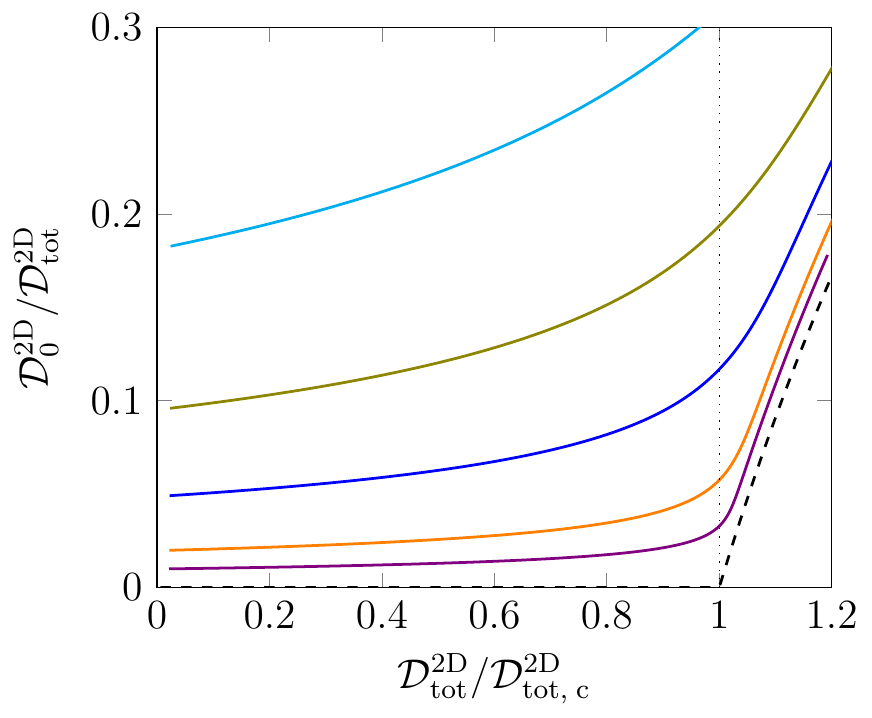}
\end{center}
\caption{\sl Variation of the (transverse) condensed fraction $f_0={\cal D}^{\rm 2D}_0/{\cal D}^{\rm 2D}_{\rm tot}$ as a function of the 
reduced total phase-space-density $x={\cal D}^{\rm 2D}_{\rm tot}/{\cal D}^{\rm 2D}_{\rm tot,\;c}$. Continuous lines from top to bottom, $\zeta=5,10,20,50,100$. In the limit $\zeta \to \infty$ (dashed black line), the crossover around $x=1$ becomes a cusp.
}
\label{fig:phase_transition}
\end{figure}

\section{Central 3D density and transverse condensation}

At thermal equilibrium the 3D density at a point $z$ is given by
\begin{equation}
\rho_{\rm tot}^{\rm 3D} (z)=\frac{1}{L^2 a_z} \sum_{\bs j} |\chi_{j_z}(z/a_z)|^2 \; n_{\bs j},
\label{}
\end{equation}
where $\chi_j$ is the $j$-th normalized Hermite function and $a_z=(1/2\pi)\,\sqrt{h/m\nu_z}$. In the limit $\lT \ll L$, we replace again the sum over $j_x,j_y$ by an integral and get
\begin{equation}
\rho_{\rm tot}^{\rm 3D} (z)=-\frac{1}{\lT^2 a_z} \sum_{ j_z} |\chi_{j_z}(z/a_z)|^2 \;\ln\left(1-Z\E^{-j_z/\zeta}\right).
\label{}
\end{equation}
In the following we are interested in the value of the 3D density in the central plane $z=0$ where it is maximum. We recall that
\begin{equation}
|\chi_{2n}(0)|^2=\frac{(2n)!}{[2^n\;n!]^2}\;\frac{1}{\sqrt \pi},\qquad \chi_{2n+1}(0)=0.
\label{}
\end{equation}
The value of the 3D phase-space-density in the plane $z=0$ is thus:
\begin{equation}
{\cal D}_{\rm tot}^{\rm 3D} (z=0)\equiv \rho_{\rm tot}^{\rm 3D}(z=0)\; \lT^3= -\sqrt{\frac{2}{\zeta}}\; \sum_{n=0}^{+\infty} \frac{(2n)!}{[2^n\;n!]^2}\;\ln\left(1-Z\E^{-2n/\zeta}  \right),
\label{eq:transverseBEC2}
\end{equation}
which is plotted in Fig.\ref{fig:3Dtot_vs_2Dtot}, as a function of ${\cal D}^{\rm 2D}_{\rm tot}$.

 \begin{figure}[t]
\begin{center}
\includegraphics{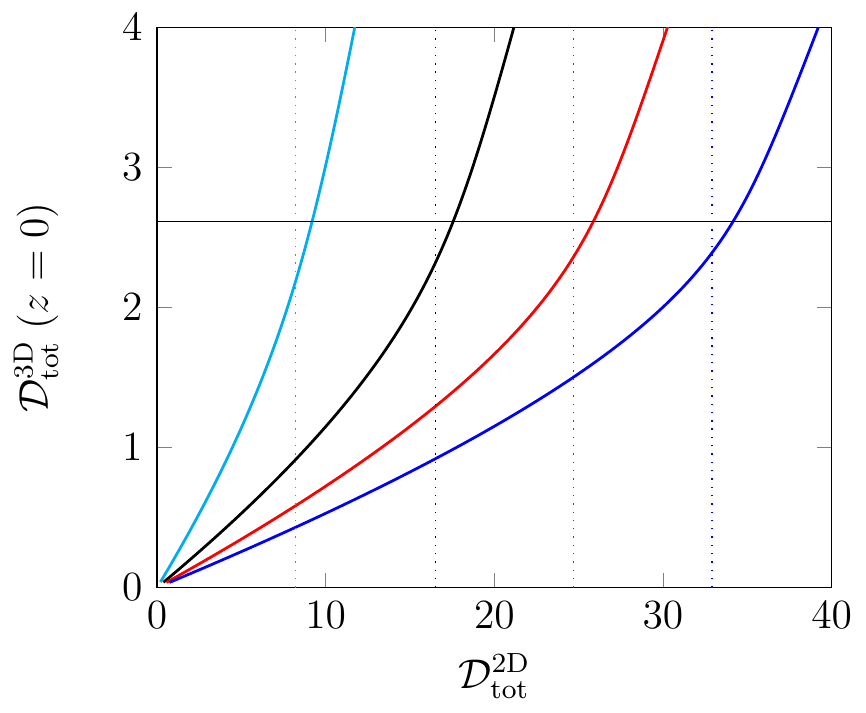}
\end{center}
\caption{\sl Total 3D phase-space-density in the plane $z=0$, as a function of the total 2D phase-space-density (same color code as in Fig.\;\ref{fig:2D_PSDs}). The horizontal line corresponds to ${\cal D}_{\rm tot}^{\rm 3D}(z=0)=2.612$, which corresponds to the threshold for BEC in a uniform 3D gas.  }
\label{fig:3Dtot_vs_2Dtot}
\end{figure}

As above we calculate the contribution of the excited states of the $z$ motion by replacing the sum over $j_z=2n$ by an integral. Using for $n\gg 1$
\begin{equation}
\frac{(2n)!}{[2^n\;n!]^2} \approx \frac{1}{\sqrt{\pi n}}
\label{}
\end{equation}
 we obtain
\begin{equation}
{\cal D}_{\rm exc.}^{\rm 3D} \approx g_{3/2}(Z),
\label{eq:transverseBEC3}
\end{equation}
Above the transverse condensation threshold, the contribution of the $j_z=0$ must be calculated separately and the total phase-space-density reads
\begin{equation}
{\cal D}_{\rm tot}^{\rm 3D} = {\cal D}^{\rm 3D}_0 + {\cal D}^{\rm 3D}_{\rm exc.}
\label{}
\end{equation}
with
\begin{equation}
\mbox{for }Z \approx 1:\qquad {\cal D}^{\rm 3D}_0 =-\sqrt{2/\zeta}\;\ln(1-Z), \qquad {\cal D}^{\rm 3D}_{\rm exc}\approx g_{3/2}(1)=2.612 ,
\label{}
\end{equation}
whereas ${\cal D}^{\rm 3D}_0 \ll {\cal D}_{\rm tot}^{\rm 3D}$ and $ {\cal D}_{\rm tot}^{\rm 3D} \approx {\cal D}^{\rm 3D}_{\rm exc}\approx g_{3/2}(Z)$ below the threshold for $\BECp$ if $\zeta \gg 1$. In other words, transverse condensation occurs when the central 3D phase-space  density approaches the value $ 2.612$, which would corresponds to a ``true" Bose--Einstein condensation for a 3D gas (Fig.\;\ref{fig:3Dtot_vs_2Dtot}).

\section{Coherence length in the $xy$ plane}

So far we have only addressed the thermodynamics of the atoms along the $z$ axis. However  when $\BECp$ occurs for $\zeta \gg 1$, the coherence length $\ell$ in the $xy$ plane  is also affected and it can become significantly larger than $\lT$. In order to characterize this in-plane coherence, we consider the one-body correlation function 
\begin{equation}
G_1(\bs r)= \langle \hat \psi^\dagger (\bs r) \, \hat \psi(0)\rangle
\label{}
\end{equation}
for $\bs r$ in the $xy$ plane. At thermal equilibrium, this is equal to
\begin{eqnarray}
G_1(\bs r)&=&\frac{1}{L^2 a_z} \sum_{\bs k, j_z} n_{\bs k, j_z}\, \E^{\I \bs k\cdot \bs r}\, |\chi_{j_z}(0)|^2 \nonumber \\
&=& \frac{1}{4\pi^2 a_z} \sum_{j_z=0}^{\infty} |\chi_{j_z}(0)|^2 \int n_{\bs k, j_z}\,\E^{\I \bs k\cdot \bs r} \;\D^2 k.
\label{}
\end{eqnarray}
Note that in this section we turn to periodic boundary conditions in order to simplify the calculations. Thus we label the single particle states by their in-plane momentum $\bs k=(2 \pi/L)(j_x,j_y)$, with $j_x,j_y \in \ZZ$. The integral over $\bs k$ can be calculated by expanding $n_{\bs k, j_z}$ as a power series in the fugacity $Z$
\begin{equation}
G_1(r) = \frac{1}{\lT^2 a_z} \sum_{j_z=0}^{\infty} \sum_{n=1}^{\infty} \frac{Z^n}{n}\; |\chi_{j_z}(0)|^2
\;\E^{-\pi r^2/n\lT^2}\;\E^{-nj_z/\zeta},
\label{eq:g1_exact}
\end{equation}
which is an infinite sum of Gaussian functions with a width increasing with $n$. 
For large $r$, it is useful to determine the dominant term in this sum over $n$, which for a given $j_z$, is obtained approximately for
\begin{equation}
n_{\rm d}(j_z)=\left( \frac{\pi r^2/\lT^2}{ j_z/\zeta\; +\;\ln(1/Z)} \right)^{1/2},
\label{}
\end{equation}
and takes a value approximately proportional to 
\begin{equation}
\frac{1}{\lT^2 a_z} \;|\chi_{j_z}(0)|^2\;\exp\left\{ -\sqrt {4\pi}\, \frac{r}{\lT}\, \left[ { j_z/\zeta\; +\;\ln(1/Z)} \right]^{1/2}  \right\}.
\label{}
\end{equation}
Let us focus on the ground state of the $z$ motion $j_z=0$ and consider the regime where $Z \approx 1$ so that $ {\cal D}^{\rm 2D}_0=-\ln(1-Z) \gg 1$ and $\ln(1/Z) \approx \E^{- {\cal D}^{\rm 2D}_0}$. We then find the approximate value  
\begin{equation}
\mbox{contribution of }j_z=0\ : \qquad G_1(r) \propto \E^{-r/\ell}, \qquad \ell=\frac{\lT}{\sqrt{4\pi}}\; \E^{{\cal D}^{\rm 2D}_0/2},
\label{}
\end{equation}  
which coincides with the result derived by another method in \cite{Hadzibabic:2011}.

A rough estimate for $\ell$ at the transverse condensation point  can be obtained by plugging the approximate value ${\cal D}^{2D}_0 \sim \ln\zeta$ obtained at Eq.\;(\ref{eq:psd0_c}) into this expression:  
\begin{equation}
\ell \approx \lT \sqrt{\zeta/ 4 \pi} = a_{\rm ho}/\sqrt 2,
\label{eq:ell_at_BECp}
\end{equation}
where $a_{\rm ho}=(2\pi)^{-1}\,\sqrt{h /m\nu_z}$. 
A more precise estimate is given in Table 1 for the relevant range of values for $\zeta$. To obtain these results, we computed numerically the variations of $G_1$ using (\ref{eq:g1_exact}), and we  looked for the wings of this function. More precisely we considered the point $r_{20}$ where the function $G_1$ is divided by 20 with respect to its value in $r=0$. At this point we define $\ell$ as 
\begin{equation}
\frac{1}{\ell} = -\left. \frac{\D \ln [G_1(r)]}{\D r}\right|_{r=r_{20}}.
\label{eq:lcoh}
\end{equation} 
If $G_1$ had an exponential variation for all $r$, this quantity would take the same value independently of the location $r$ where it is calculated. For a non-strictly exponential $G_1$, the present definition is a good compromise between considering the far wings of $G_1$ in order to monitor the appearance of an extended coherence, and restricting to sufficiently small values of $r$ so that the values of $G_1$ are still significant. The variation of $\ell/\lT$ with ${\cal D}_{\rm tot}^{\rm 2D}$ for various values of $\zeta$ is shown in Fig.\;\ref{fig:lcoh_vs_3Dtot}. 

\begin{figure}[t]
\begin{center}
\includegraphics{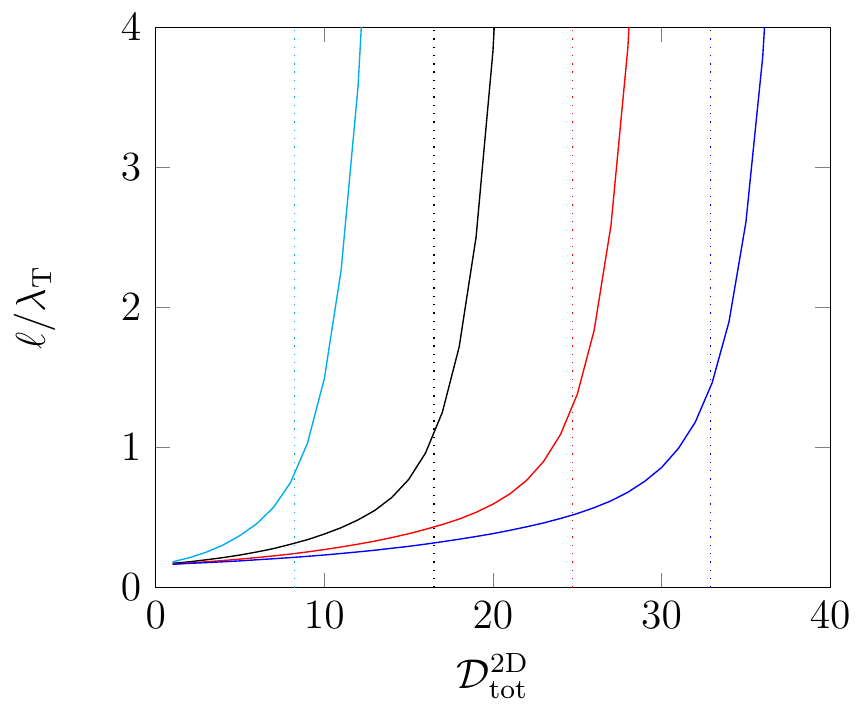}
\end{center}
\caption{\sl Variations of the coherence length $\ell$ defined in (\ref{eq:lcoh}) with the total 2D phase-space-density (same color code as in Fig.\;\ref{fig:2D_PSDs}).}
\label{fig:lcoh_vs_3Dtot}
\end{figure}

For Boltzmann statistics $G_1(r)=\exp(-\pi r^2/\lT^2)$, so that the expected value of $\ell$ at $r=r_{20}$ is 
$\ell^{\rm Bolt.}=0.16\,\lT$. We see in table \ref{tab:typical} that at \BECp, the value of $\ell$ is increased by a factor $6-9$  for $\zeta=10-20$ with respect to $\ell^{\rm Bolt.}$.
More precisely the result (\ref{eq:ell_at_BECp}) states that at the threshold for \BECp, the coherence length in the $xy$ plane is comparable to the size of the ground state along the $z$ direction. The physical meaning of (\ref{eq:ell_at_BECp}) is related to the fact that the 3D phase-space-density in the plane $z=0$ reaches the value $2.612$ when \BECp\, occurs. For a uniform, infinite 3D gas with this spatial density, the coherence length $\ell$  would diverge, signaling the occurrence of a true Bose--Einstein condensation. Here the $z$ confinement  limits the extension of the coherent part of the gas along $z$ to $a_{\rm ho}$, which prevents the divergence  of $\ell$ and limits its value also to $a_{\rm ho}$ in the $xy$ plane. In the regime $\zeta \gg 1$, the appearance of a large coherence length in the $xy$ plane and the occurrence of transverse condensation for the $z$ degree of freedom are thus linked. 

\section{Finite size effects}

So far we assumed that the size of the box in the $xy$ plane was arbitrarily large compared to the other length scales of the problem, $\lT$ in particular, so that the sum over $j_x,j_y$ could safely be replaced by an integral. For a finite-size box this approximation ceases to be valid when the phase-space-density becomes large enough. A full 3D Bose--Einstein condensation (\BECf) can then take place, with a macroscopic accumulation of particles in the single-particle ground state $j_x=j_y=1$, $j_z=0$. The existence of two successive condensations when the phase-space-density increases, first \BECp \ and then \BECf,  was highlighted in \cite{vand97}.

 \BECf\  is expected to occur when the chemical potential is chosen smaller (in absolute value) than the gap $3 h^2/(8mL^2)$ between the true ground state of the box and the first excited states. At the point where \BECf\ occurs, the 2D phase-space-density associated to the $|j_z=0\rangle$ level is
\begin{equation}
 {\cal D}^{\rm 2D}_0 =-\ln(1-Z) \approx \ln \left[{\kB T}/{|\mu|}\right] \approx \ln \left[(4/3\pi)\,L^2/\lT^2\right].
\label{BECf}
\end{equation}
 This value is notably larger than the value  $ {\cal D}^{\rm 2D}_0=\ln(\zeta)$ when $\BECp$ occurs if
\begin{equation}
\frac{\zeta}{(4/3\pi)\,L^2/\lT^2} =\frac{3\pi^2}{2}\; \frac{a_z^2}{L^2} \ll 1,
\label{}
\end{equation}
meaning simply that the gas must have a flat shape.

An example is given in Fig.\;\ref{fig:two_transitions} where we plot the result of a calculation summing the populations of the individual quantum states given in (\ref{eq:quantum_states}). The calculation is performed for a typical box size $L=200\,\lT$ and for $\zeta=5$. The two successive transitions are clearly visible on this figure. First when $ {\cal D}^{\rm 2D}_{\rm tot} \approx  {\cal D}^{\rm 2D}_{\rm tot,\ c}$ with ${\cal D}^{\rm 2D}_{\rm tot,\ c}=\zeta \pi^2/6 \approx 8.2$, the phase-space-density associated to the excited state of the $z$ motion saturates at a value close\footnote{The asymptotic value is not exactly $\zeta\pi^2/6$, because $\zeta$ is not very large compared to 1 in this example.} to ${\cal D}^{\rm 2D}_{\rm tot,\ c}$, which is the signature of $\BECp$. Then when ${\cal D}^{\rm 2D}_{\rm tot}$ reaches a value around
\begin{equation}
{\cal D}^{\rm 2D}_{\rm tot}=\zeta \pi^2/6 + \ln \left[(4/3\pi)\,L^2/\lT^2\right] \approx 18,
\label{eq:BECfull}
\end{equation}
the fraction of atoms $f_{\rm BEC}$ occupying the  overall ground state starts to be significant. It is clear on this figure that there exists a domain of values of ${\cal D}^{\rm 2D}_{\rm tot}$ for which the population of the excited states of the $z$ motion is saturated, with no macroscopic occupation of the single particle ground state in the box.

\begin{figure}[t]
\begin{center}
\includegraphics{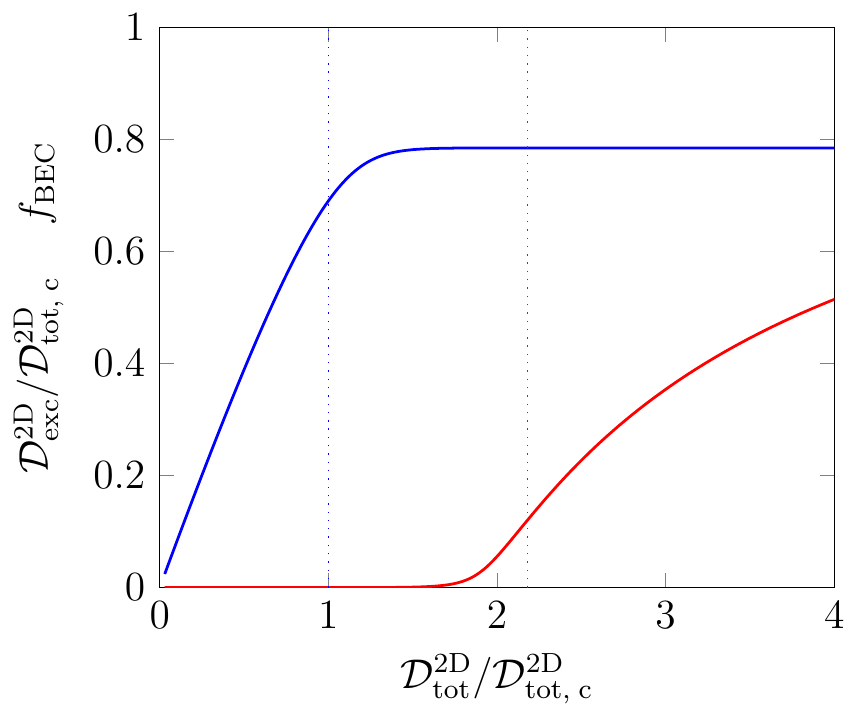}
\end{center}
\caption{\sl Continuous blue line: phase-space-density associated to the atoms occupying the excited states of the $z$ motion. Continuous red line: Fraction of atoms in the ground state in the box. The vertical dotted lines give the positions of the expected transitions \BECp\ (blue, Eq.\;\,\ref{eq:BECp}) and \BECf\ (red, Eq.\,\ref{eq:BECfull}).  Calculation performed for $L=200\lT$ and $\zeta=5$. }
\label{fig:two_transitions}
\end{figure}

For this typical box size $L=200\lT$, we infer from Eq.\;(\ref{BECf}) that $\BECf$\ occurs when $ {\cal D}^{\rm 2D}_0$ reaches the value $\sim 9.7$. For Rb atoms and a trapping frequency $\nu_z$ in the range $300-1500$\, Hz, this phase-space-density is similar to the one at which the (interaction-induced) BKT superfluid transition occurs (${\cal D}^{\rm 2D}_{\rm BKT}=8-10$) \cite{Prokofev:2001,Hadzibabic:2011}. Therefore the relevance of the previous sections of this supplementary material, based on calculations for a non-interacting gas in the limit $L \gg \lT$ (no $\BECf$), is limited to the region of parameters where $ {\cal D}^{\rm 2D}_0 \lesssim 8$. This is precisely the region where the emergence of coherence studied in the paper occurs ( $ {\cal D}^{\rm 2D}_0 \sim 4$).

\end{document}